\documentclass[twocolumn,showpacs,amsmath,amssymb,pra,longbibliography
]{revtex4-1}

\usepackage[pdftex]{graphicx}
\usepackage[usenames,dvipsnames]{color}
\usepackage{epstopdf}
\usepackage[hidelinks]{hyperref}
\usepackage{tabularx}

\newcommand{\p}[1]{\mathbf{p}}
\newcommand{\q}[1]{\mathbf{q}}

\newcommand{\PP}[1]{\mathrm{(L)}}
\newcommand{\AP}[1]{\mathrm{(R)}}

\newcommand{\w}[1]{\mathbf{\hat w}}

\newcommand{\z}[1]{\theta}
\newcommand{\mix}[1]{\Omega}

\newcommand{\aop}[1]{\hat{a}}
\newcommand{\cop}[1]{\hat{c}}

\newcommand{\sm}[1]{\mathrm{SM}}
\usepackage{slashed}

\usepackage{dcolumn}
\newcolumntype{d}[1]{D{.}{.}{#1}}

\newcommand{\U}[1]{\mathbf{U}{}}
\newcommand{\D}[1]{\mathbf{D}{}}
\newcommand{\V}[1]{\mathbf{V}{}}

\usepackage{booktabs}
\usepackage{siunitx}

\usepackage[T1]{fontenc}
\usepackage{beramono}
\usepackage{listings}

\lstdefinelanguage{Julia}%
  {morekeywords={abstract,break,case,catch,const,continue,do,else,elseif,%
      end,export,false,for,function,immutable,import,importall,if,in,%
      macro,module,otherwise,quote,return,switch,true,try,type,typealias,%
      using,while},%
   sensitive=true,%
   alsoother={$},%
   morecomment=[l]\#,%
   morecomment=[n]{\#=}{=\#},%
   morestring=[s]{"}{"},%
   morestring=[m]{'}{'},%
}[keywords,comments,strings]%

\usepackage{graphicx}
\makeatletter
\newcommand*\bigcdot{\mathpalette\bigcdot@{1}}
\newcommand*\bigcdot@[2]{\mathbin{\vcenter{\hbox{\scalebox{#2}{$\m@th#1\bullet$}}}}}
\makeatother

\graphicspath{{FIGs/}}

    \makeatletter
\def\@fnsymbol#1{\ensuremath{\ifcase#1\or \dagger\or \ddagger\or 
   \mathsection\or \mathparagraph\or \|\or **\or \dagger\dagger
   \or \ddagger\ddagger \else\@ctrerr\fi}}
    \makeatother

\begin{document}

\title{Direct solution of multiple excitations in a matrix product state with block Lanczos}
\author{Thomas E.~Baker}
\affiliation{Institut quantique \& D\'epartement de physique, Universit\'e de Sherbrooke, Sherbrooke, Qu\'ebec J1K 2R1 Canada}
\affiliation{Department of Physics, University of York, Heslington,York YO10 5DD, United Kingdom}
\affiliation{Department of Physics \& Astronomy; Department of Chemistry; Centre for Advanced Materials and Related Technologies, University of Victoria, Victoria, British Columbia V8P 5C2, Canada}
\author{Alexandre Foley}
\affiliation{Institut quantique \& D\'epartement de physique, Universit\'e de Sherbrooke, Sherbrooke, Qu\'ebec J1K 2R1 Canada}
\author{David S\'en\'echal}
\affiliation{Institut quantique \& D\'epartement de physique, Universit\'e de Sherbrooke, Sherbrooke, Qu\'ebec J1K 2R1 Canada}

\date{\today}

\begin{abstract}

Matrix product state methods are known to be efficient for computing ground states of local, gapped Hamiltonians, particularly in one dimension.  We introduce the multi-targeted density matrix renormalization group method that acts on a bundled matrix product state, holding many excitations.  The use of a block or banded Lanczos algorithm allows for the simultaneous, variational optimization of the bundle of excitations.  The method is demonstrated on a Heisenberg model and other cases of interest.  A large of number of excitations can be obtained at a small bond dimension with highly reliable local observables throughout the chain.

\end{abstract}

\maketitle


\section{Introduction}

The density matrix renormalization group (DMRG) has become one of the premier methods to solve lattice Hamiltonians \cite{white1992density,white1993density,schollwock2005density,schollwock2011density}.  DMRG breaks the quantum problem apart into individual tensors, one representing each site on the lattice. The renormalization of the full problem to a reduced number of degrees of freedom allows for high accuracy but with dramatically reduced memory.  Relying on the properties of entanglement, DMRG is capable of finding the ground state for lattice Hamiltonians quickly and efficiently.  Systems in higher dimensions (greater than one dimension) are harder to converge, but many are still reachable \cite{stoudenmire2012studying}.

Soon after DMRG's derivation, the matrix product state (MPS) was identified as the underlying wavefunction ansatz \cite{ostlund1995thermodynamic,rommer1997class,affleck1988valence}. This provides a useful way to represent the tensors of the full quantum problem. An interesting feature of the MPS formalism is that a single tensor, called the orthogonality center, contains all information that is passed between two partitions of the system.  It can then be said the MPS is gauged in a certain way, depending on where the center of orthogonality is located.  Because only one tensor contains the weights of the density matrix, this makes local measurements very efficient.  The MPS efficiently represents the ground state of gapped systems. By truncating the eigenvalues of the density matrix, an accurate representation of the full, exponentially sized ground state can be obtained with this method.  The ability to truncate the MPS is one of the reasons that DMRG scales far better than exact diagonalization without sacrificing too much accuracy.  In comparison with quantum Monte Carlo, the DMRG algorithm does not suffer a sign problem \cite{feldt2020excited}.

Similarly to the MPS representation, the matrix product operator (MPO) is a site-by-site representation of operators in the system. Defining an MPO allows for the use of the DMRG algorithm to obtain the ground state solution using only a few sites in the system instead of using an operator that is applied on all sites at once.

Developing algorithms for the computation of excited states has received significant effort \cite{vanderstraeten2019simulating,d2019targeting,ponsioen2020excitations,boschi2004investigation}.  Even in the original DMRG paper, the topic of excitations are discussed \cite{white1993density}, and excitations for single frequencies for Green's functions were investigated with traditional Lanczos recursion \cite{hallberg1995density,jeckelmann2002dynamical}.  

The most widely used method of solving for excitations in tensor network involve adding a penalty term to the Hamiltonian, $H_0$.  The procedure is iterative and involves solving a Hamiltonian $H_0$ for the ground state, $|\psi_0\rangle$.  The Hamiltonian is modified as
\begin{equation}
H=H_0+\lambda|\psi_0\rangle\langle\psi_0|
\end{equation}
where a scalar quantity $\lambda$ is added to effectively modify the energy by $E_0+\lambda$.  Thus, by adding a penalty term, the energy of the ground state is shifted.  If the variational minimum of $H$ is now sought, then the minimum energy will now be the first excited state of $H_0$. The process can be repeated by adding an energetic penalty for the excited states, leading to a solution corresponding to another excited state. This method grows the bond dimension and can have trouble converging to the excited states with the typical DMRG solution.

There are also other strategies to find excitations. In one very interesting result \cite{chepiga2017excitation}, the solution of many excitations were obtained from a Lanczos recursion on a single-site of the network. This was demonstrated to give high quality excitations that eventually decayed with more iterations due to numerical roundoff errors native to Lanczos methods.  The method also performed best in the middle of the MPS where the local Hilbert space was largest. The ideas from this paper recently led to another post-processing algorithm that derived excitations from the ground state by finding the null space of the tensor representation \cite{van2021efficient}.  With recent interest in many-body localization, still other methods have attempted to find highly excited states \cite{khemani2016obtaining,yu2017finding}.

While methods that rely on a postprocessing step may be fast and accurate in some cases, it should be expected that inaccuracies in correlation functions and energies will appear for higher energy excitations. This is evident in Ref.~\onlinecite{chepiga2017excitation} at higher energy excitations and will be present in similar post-processing methods.  In particular, at the edges of a chain, where the bond dimension tapers off from its bulk value to a value of 1 \cite{bakerCJP21,*baker2019m,schollwock2005density}, the size of the local Hilbert space also becomes smaller. Thus, when using a Lanczos method here, the decay in the accuracy of the Lanczos coefficients is even more dramatic. Post-processing methods to find excitations are ultimately limited by the size of the Hilbert space on which the necessary Lanczos operations are performed.  When the bond dimension is large, the effects will be reduced as in, for example, a critical system, for long range interaction, or periodic bonds.  Finding a method to find excitations away from this constrained class of methods would be useful. 

One other way to obtain excitation is to optimize a bundle of states.   The strategy used in Ref.~\onlinecite{huang2018generalized} is to attach a single index to the orthogonality center which labels the wavefunctions.  This effectively bundles states together to make a bundled MPS. However, Ref.~\onlinecite{huang2018generalized}'s algorithm optimizes a generic cost function.  From the bundled MPS, the matrix blocks of the Hamiltonian necessary for block Lanczos can be recovered as a last step in that algorithm, but they were not directly solved. 

This is where the ideas presented in this paper depart from the previous literature.  The algorithm that is presented here effectively replaces the Lanczos update step in DMRG with a block Lanczos (or banded Lanczos or similar) algorithm.  Block and banded Lanczos methods extend the traditional use of Lanczos from scalar coefficients to matrix coefficients \cite{cullum1974block,bai2000templates,senechal2008introduction}. The implementation of the block Lanczos algorithm is shown here to resolve degeneracies to a high accuracy and has already solved large scale systems with no detectable issue provided that a sufficiently high bond dimension is used \cite{di2021efficient}. Additionally, there is little extra cost to the size of the wavefunction.  Many tens or hundreds of excitations have been solved in the models we investigate with only a moderately large bond dimension.  The advantage of directly infusing the DMRG algorithm with block Lanczos will avoid the issues with solving a generic cost function to obtain the excitations \cite{huang2018generalized,khemani2016obtaining,yu2017finding}.

To debut this multi-targeted DMRG method, first the concept of a bundled MPS is introduced formally in Sec.~\ref{bundleMPS}.  Then, the block Lanczos algorithm as a tensor network routine is introduced.  In Sec.~\ref{multiDMRG}, the multi-targeted DMRG algorithm is introduced for both dense tensors and quantum-number tensors.  A single-site and a two-site algorithm are presented, although it is demonstrated that the single-site algorithm is sufficient and has no disadvantage over the slower two-site algorithm.  Sec.~\ref{models} tests the model to find fast and accurate solutions in comparison with exact diagonalization.  This also works for degeneracies and even works on small bond dimension models.

\section{Matrix Product States of excitations}

The basic notation and diagrams are reviewed in Appendix~\ref{backgroundinfo}, which introduces briefly the DMRG algorithm, matrix product states, and some considerations that are necessary here.

\begin{figure*}
\includegraphics[width=1.25\columnwidth]{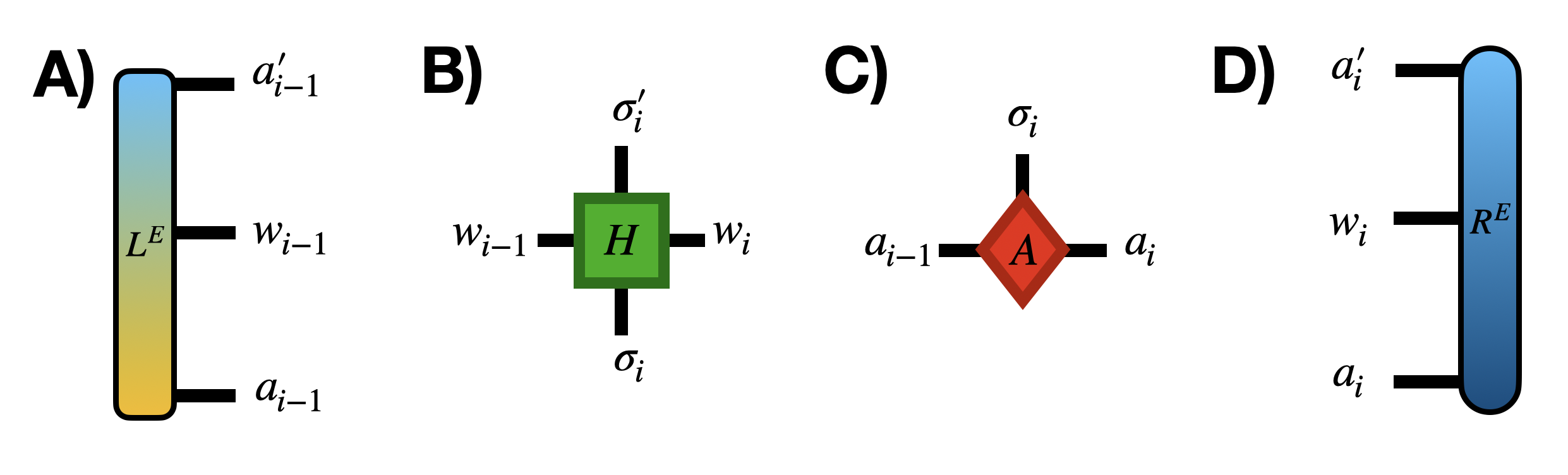}
\caption{Tensor indices are shown on the diagrams common to tensors in the A) left-environment, $(L^E)_{a'_{i-1}w_{i-1}a_{i-1}}$, B) MPO, $H_{w_{i-1}w_i}^{\sigma_i\sigma_i'}$, C) MPS, $A^{\sigma_i}_{a_{i-1}a_i}$, and D) right-environment, $(R^E)_{a_i'w_ia_i}$. Raised and lowered indices are only for presentation purposes and have no additional meaning here. The color gradient chosen on the left- and right-environment tensors is for effect only and is a composite of the tensor colors used to construct those tensors.}
\label{tensorlegend}
\end{figure*}

Throughout the text, we use a graphical notation for tensor networks common in the field \cite{bakerCJP21,*baker2019m}. Each shape on the diagram represents a tensor and the legs indicate the indices on the tensor.  Normally, the legs are not labeled with tensor indices, but we do so in Fig.~\ref{tensorlegend} for readability here and in some diagrams below.

\subsection{Bundled matrix product state}\label{bundleMPS}

To describe a bundle of states on a single MPS, a new index must be added to one of the sites.  The fact that not all sites must receive another excitation is an expression of the area law \cite{dmrjulia1}.  By recording information of the entanglement between two parts of a system, the MPS naturally is expressed with a single orthogonality center.  The same line of thinking applies to a group of excitations as can be seen from the bundling of states in the following.

Attaching the extra index, $\xi$ (referred to as the excitation index throughout), to the orthogonality center, the MPS takes the form
\begin{align}\label{bundleMPS}
|\Psi\rangle=&\sum_{\substack{\{a_j\},\\ \{\sigma_j\}, \xi}}A^{\sigma_1}_{a_1}\ldots A^{\sigma_{j-1}}_{a_ja_{j-1}}
D^{\sigma_j\xi}_{a_{j-1}a_j}B^{\sigma_{j+1}}_{a_{j}a_{j+1}}\ldots B^{\sigma_N}_{a_{N-1}}\nonumber\\
&\hspace{3cm}\times|\sigma_1\sigma_2\ldots\sigma_N\rangle|\xi\rangle
\end{align}
in a mixed gauge (containing both left- and right-normalized tensors, $A$ and $B$, respectively) with the orthogonality center on site $j$. The bundled MPS is gauge invariant just as a regular MPS.

There is a way to describe this bundle of excitations as a special direct sum over specific indices of a family of MPSs.  To formalize this concept, some special notation must be introduced.  Just as with contraction and decomposition, the tensor can be permuted and reshaped into a matrix to perform the operation on the matrix-equivalent \cite{bakerCJP21,*baker2019m}. The same is true for the direct sum on two vectors.  To review, the direct sum for two vectors $v_1$ and $v_2$ results in a matrix of those vectors as
\begin{equation}
v_1\oplus v_2=\left(\begin{array}{cc}
v_1 & v_2
\end{array}
\right)
\end{equation}
Similarly, the operation can be applied on more than two vectors,
\begin{equation}\label{directsumvector}
\bigoplus_j v_j=\left(\begin{array}{cccc}
v_1 & 
v_2 &
\dots & 
v_N
\end{array}
\right)
\end{equation}
For the case of tensors, one can group all indices into two groups. One group is the index that will be expanded in size.  The other indices are kept the same.  This reduction of the tensor object to something that can be included in the direct sum of vectors allows for the use of the traditional operation.

The indices will be grouped into two groups using a tilde notation, indicating the indices that will be expanded on. For example,

\begin{equation}\label{notationsum}
A^{\sigma_i}_{a_{i-1}\tilde a_i}\oplus B^{\sigma_i}_{a_{i-1}\tilde a_i} = C^{\sigma_i}_{a_{i-1}a_i}
\end{equation}
means that a tensor $A_{a_{i-1}a_i}^{\sigma_i}$ and $B_{a_{i-1}a_i}^{\sigma_i}$ were joined along the index $a_i$.  In general, any indices with a tilde symbols are expanded and the others are not.

The direct sum operator $\oplus$ can be used to indicate how indices are joined together to form the bundled MPS \cite{bakerCJP21,*baker2019m}.  For left-normalized tensors, the expansion appears as
\begin{equation}\label{leftbundle}
\bigoplus_j \left(A_j\right)^{\sigma_i}_{\tilde a_{i-1}\tilde a_i}=A^{\sigma_i}_{a_{i-1}a_i}\quad\mathrm{[left-normalized]}
\end{equation}
and similarly for the right-normalized tensors
\begin{equation}\label{rightbundle}
\bigoplus_j \left(B_j\right)^{\sigma_i}_{\tilde a_{i-1}\tilde a_i}=B^{\sigma_i}_{a_{i-1}a_i}\quad\mathrm{[right-normalized]}
\end{equation}
where both $A$ and $B$ are taken from the individual excitations (index $j$) joined together into the bundle; however, the excitation index can be attached to any single tensor of the bundled MPS.

The orthogonality center was already identified as holding the excitations in Eq.~\eqref{bundleMPS}, so the individual tensors across each MPS must be joined together as
\begin{equation}\label{ocD}
\bigoplus_\xi \left(D^\xi\right)^{\sigma_i\bigcdot}_{\tilde a_{i-1}\tilde a_i}\rightarrow D^{\sigma_i\xi}_{a_{i-1}a_i}\quad\mathrm{[center\;of\;orth.]}.
\end{equation}
where $\xi$ indexes the individual tensors.  Further, a new symbol $\bigcdot$ is added to indicate that an index of size 1 is added onto the tensor and implicitly means that the extra index is kept as a record of which excitation is being examined. For example, Eq.~\eqref{directsumvector} would become with this new notation,
\begin{equation}
\bigoplus_j(v_j)_i^{\bigcdot}\rightarrow v_i^j
\end{equation}
where the index $j$ records each of the vectors $v$ with elements indexed by $i$. The resulting tensor $v_i^j$ can be interpreted as a matrix ($v_{ij}$) or, as was the case in Eq.~\eqref{directsumvector}, a vector of vectors. Note that none of the indices receive a tilde (as was the case in Eq.~\eqref{ocD}) because none of the indices is expanded in the direct sum.

The form in Eq.~\eqref{ocD} is the closest to the traditional use of the direct sum operator for matrices $M_j$ \cite{bakerCJP21,*baker2019m},
\begin{equation}\label{directsummatrices}
\bigoplus_j M_j = \left(\begin{array}{ccccc}
M_1 & 0 & 0 & \cdots & 0\\
0 & M_2 &0 & \ddots & 0\\
0 & 0 & M_3 & \ddots & 0\\
\vdots & \ddots & \ddots & \ddots & \vdots\\
0 & 0 & 0 &\cdots & M_N
\end{array}\right)
\end{equation}
although there is a key difference between Eq.~\eqref{directsummatrices} and Eq.~\eqref{directsumvector} in that an implicit reshaping of an index of size 1 onto the objects.  This appears in Eqs.~(\ref{leftbundle}--\ref{rightbundle}) and Eq.~\ref{ocD} since the direct sum index does not appear on the final tensor in the former and does in the latter (appearing at $\xi$). This implicitly expands the auxiliary indices as well. 

Typically, the excitation index $\xi$ will be stored on one of the unitaries instead of the orthogonality center itself.  This is a useful practice when considering the singular value decomposition (SVD) or other decomposition and deciding where to attach the excitation index.

The construction above represents the largest MPS bundle that can be formed for the excitations.  The bond dimension of this bundle is the direct sum of all the constituent wavefunctions that were used to construct it, which effectively overdetermines the basis states, and generates a lot of parallel basis functions in the tensor network.  A sweep over all sites in the system will allow the SVD or other decomposition to compress the bond dimension, but there will be a minimum bond dimension to represent the excitations without sacrificing fidelity of the wavefunction. We note that orthogonality of the left- and right-normalized tensors is maintained because Eq.~\eqref{ocD} is fully block diagonal, ensuring that orthogonality is preserved between elements of the blocks.

Note that one can recreate the form of the MPS constructed from the direct sum operations above if a single excitation is selected from the bundled MPS ansatz. Then performing the same for all $g$ MPSs representing excitations, the over-determined tensors can be recovered.

In practice, the bundle of excitations will not be required to start the algorithm.  One can simply reshape an arbitrary MPS's orthogonality center with an extra index of size one, trivially, and then begin the algorithm.  The excitations will be discovered with an additional ingredient which is introduced next. The case of symmetries can be initialized with a systematic treatment. If a set of quantum numbers is not present in the bundle at a given time, then a random tensor initialized in the proper quantum number can be created and used in the algorithm. At this point, the SVD similar to Fig.~\ref{blocklanczosfig}a can be used to generate normalized blocks when this step is triggered.

\section{Density matrix renormalization group on many excitations}\label{multiDMRG}

Simply defining the bundled MPS also requires an operation to act on all excitations at once. The task now is to identify how to optimize the states simultaneously and variationally.  The key step in the DMRG algorithm is to modify the Lanczos step so that it acts on the bundled MPS.  Instead of using the traditional Lanczos recursion to find the updated tensors, a block Lanczos technique is used \cite{cullum1974block,bai2000templates,senechal2008introduction}.  In this case, the use of block Lanczos has only to do with the choice of gauge in the resulting decomposition.  If another choice were made, then it would be possible to obtain another extended form of Lanczos, such as banded Lanczos (which would be obtained if a QR decomposition is used instead of the SVD in the following).

Block Lanczos will be presented first in its full form before applying it onto the bundle of excitations in a two-site and single-site algorithm.

\subsection{Block Lanczos}\label{blocklanczos}

The traditional Lanczos algorithm is a recursion relation where a linear combination of previously obtained states are multiplied with scalar coefficients $\alpha$ and $\beta$.  The recursion relation is given in Appendix in Eq.~\eqref{lanczos}.  The advantage of this method over a power method is that fewer operations are required to generate the ground state wavefunction.

Block Lanczos is an extension of the Lanczos algorithm that generates the Krylov subspace of a set of states instead of a single one~\cite{cullum1974block}.
In the context of DMRG, it is perfectly suited to updating the orthogonality center of a bundled MPS.

The recurrence relation of block Lanczos is 
\begin{equation}
	\label{eq:BL-rec}
	\Psi_{n+1}\mathbf{B}_{n+1} = H\Psi_n - \Psi_n\mathbf{A}_n-\Psi_{n-1}\mathbf{B}_{n}^\dagger
\end{equation}
where the $\Psi_n$ are rectangular $K\times g$ matrices composed of mutually orthonormal columns. The column length $K$ is the size of the Hilbert space of $H$ and the number of columns $g$ is the block dimensions, a parameter of the algorithm. The $\mathbf{A}_n$ and $\mathbf{B}_n$ are $g\times g$ matrices.

The matrices in Eq.~\eqref{eq:BL-rec} are defined by the following equations:
\begin{align}
	\label{eq:Psiortho}
	\Psi_n^\dagger\Psi_k &= \mathbb{I}_{g\times g} \delta_{nk} \mathrm{,}\\
	\label{eq:defA}
	\mathbf{A}_n &= \Psi_n^\dagger H \Psi_n \mathrm{,}\\
	\label{eq:defB}
	\mathbf{B}_n &= \Psi_n^\dagger H \Psi_{n-1} \mathrm{.}
\end{align}
where $\mathbb{I}_{g\times g}$ is a size $g$ identity matrix and $\delta_{mn}$ is Kronecker's delta.  
The recurrence relation~Eq.~\eqref{eq:BL-rec} has the following initial conditions: $\Psi_{-1} = 0$ and $\Psi_0$ is an arbitrary matrix satisfying Eq.~\eqref{eq:Psiortho}. The columns of $\Psi_0$ are the states for which the block Lanczos algorithm generates the Krylov subspace. Note that by choosing $g=1$, we recover the usual Lanczos algorithm. In that case, the matrices $\mathbf{A}_n$ and $\mathbf{B}_n$ are reduced to the $\alpha_n$ and $\beta_n$ coefficients. Note that the $\Psi_n$ matrices can be related to the notation for the scalar Lanczos case (see Eq.~\eqref{lanczos}) as
\begin{equation}\label{excitationvec}
\Psi_n=\left(\begin{array}{ccccc}
|\psi_1^{(n)}\rangle&|\psi_2^{(n)}\rangle&|\psi_3^{(n)}\rangle&\ldots&|\psi_g^{(n)}\rangle
\end{array}\right)
\end{equation}
where each of the $|\psi_i^{(n)}\rangle$ is a vector of size $N$ and can be placed as a column in the $\Psi_n$ matrix.

In practice, $\mathbf{A}_n$ can be computed from its definition~Eq.~\eqref{eq:defA}, but $\Psi_n$ and $\mathbf{B}_n$ cannot be computed from their respective definitions. Instead one must proceed as follows: Given the quantities resulting from the previous iteration, we can compute
\begin{equation}
	\widetilde{\Psi}_{n+1} =
	 \Psi_{n+1}\mathbf{B}_{n+1} =
	  H\Psi_n 
	  - \Psi_n\mathbf{A}_n
	  - \Psi_{n-1}\mathbf{B}_n^\dagger \mathrm{.} \label{eq:pract_iter}
\end{equation}
To obtain $\Psi_{n+1}$, we must orthonormalize the columns of $\widetilde{\Psi}_{n+1}$, the (non-unitary) matrix transform that accomplish this goal is $\mathbf{B}_{n+1}^{-1}$.
We proceed by performing a matrix decomposition such as the SVD or the QR decomposition. The choice of decomposition is a choice of gauge, with no fundamental consequences. We choose to use the SVD in this way:
\begin{align}\label{nextPsi}
	\mathbf{U}_{n+1} \mathbf{D}_{n+1} \mathbf{V}_{n+1}^\dagger = & \widetilde{\Psi}_{n+1} \mathrm{,} \\
	\mathbf {U}_{n+1} \mathbf{V}_{n+1}^\dagger \equiv & \Psi_{n+1} 
	\mathrm{,}\\
	\mathbf{V}_{n+1} \mathbf{D}_{n+1} \mathbf{V}_{n+1}^\dagger = & \mathbf{B}_{n+1} =  \mathbf{B}^\dagger_{n+1} \mathrm{.}
\end{align}
where it is noted that with this choice of gauge, $\mathbf{B}_n^\dagger = \mathbf{B}_n$. To obtain $\Psi_{n+1}$ and $\mathbf{B}_{n+1}$, a closure relationship was inserted into the decomposition of $\tilde\Psi_{n+1}$. The form chosen here is the polar decomposition \cite{bhatia1997matrix}; however, any unitary matrix can be inserted into the SVD and give another form without changing the overall properties of the system.  The QR decompositions could be used here but would give a different form for the matrices, although when implemented the results presented in the following do not change. 

With these practical definitions in hand, one can easily verify that we satisfy Eq.~\eqref{eq:pract_iter}. Assuming the non-singularity of the $\mathbf{B}_n$ one can demonstrate by induction that Eqs.~\eqref{eq:Psiortho} and~\eqref{eq:defB} are satisfied.

Once we have performed $p$ steps of the block Lanczos recursion, the Hamiltonian takes a block tridiagonal form in the orthonormal basis defined by the columns of the $\Psi_n$ matrices:
\begin{align}
	\mathbf{T}_p &= (\Psi_0,\Psi_1,\dots,\Psi_p) \label{vectorbundle}\\
	\check{\mathbf{M}} \equiv \mathbf{T}_p^\dagger H\mathbf{T}_p &=\left(\begin{array}{ccccc}
		\mathbf{A}_{0} & \mathbf{B}_{1}^{\dagger} & \mathbf{0} & \cdots & \mathbf{0} \\
		\mathbf{~B}_{1} & \mathbf{~A}_{1} & \mathbf{~B}_{2}^{\dagger} & \cdots & \mathbf{0} \\
		\mathbf{0} & \mathbf{~B}_{2} & \mathbf{~A}_{2} & \ddots & \vdots \\
		\vdots & \vdots & \ddots & \ddots & \mathbf{B}_{n}^{\dagger} \\
		\mathbf{0} & \mathbf{0} & \cdots & \mathbf{B}_{n} & \mathbf{A}_{n}
		\end{array}\right)\label{blockM}
\end{align}
where it should be noted again that the arrangement of bundled states in Eq.~\eqref{vectorbundle} can be represented as a matrix of size $K\times (pg)$. 

This block diagonal matrix can be diagonalized and its lowest eigenvalues are reasonable approximation of the lowest eigenvalues of $H$. Degeneracies of degree $g$ or less are resolved. In the context of DMRG, one step of this algorithm allows us to update the $g$ eigenstates candidates contained in the orthogonality center of the bundled MPS.

\subsection{Block Lanczos on a bundled MPS tensor}

\begin{figure}
\includegraphics[width=\columnwidth]{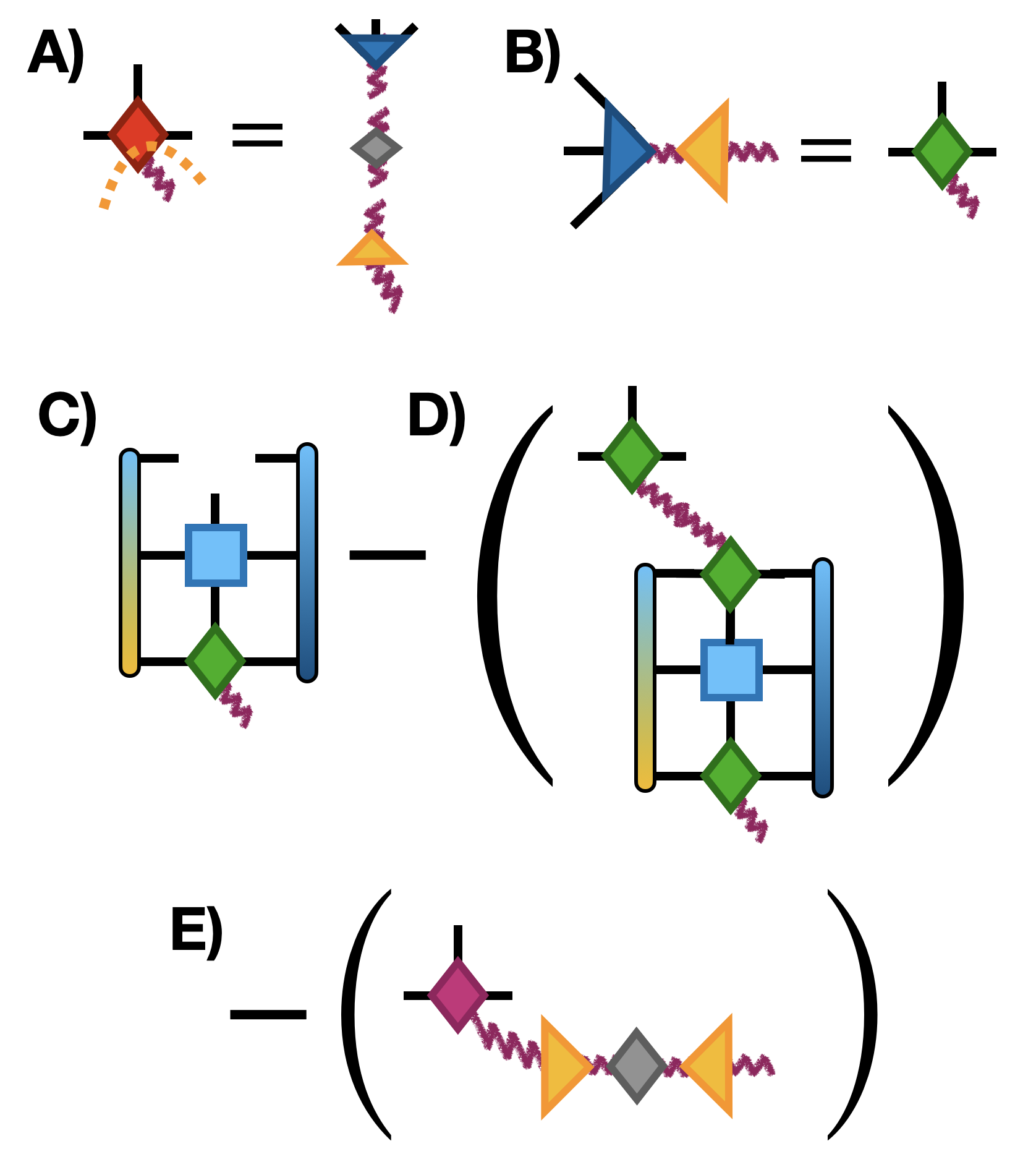}
\caption{The block Lanczos recursion relation formulated in diagrams corresponding to Eq.~\eqref{eq:BL-rec}. The color of the gradients for the environment tensors are chosen arbitrarily to correspond to the left- and right-normalized tensors in the network.  The meaning of each tensor in the diagram is explained in the text. The new indices introduced by the SVD are in the same basis as the excitation index but are shown with a solid line for clarity in the figure.}
\label{blocklanczosfig}
\end{figure}

\begin{figure}[b]
\begin{center}
\includegraphics[width=0.8\columnwidth]{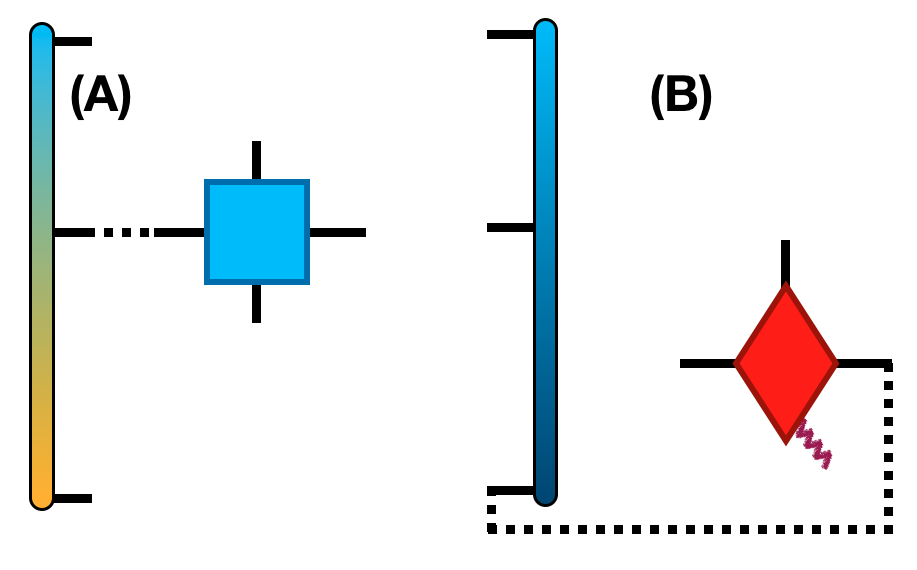}
\includegraphics[width=0.4\columnwidth]{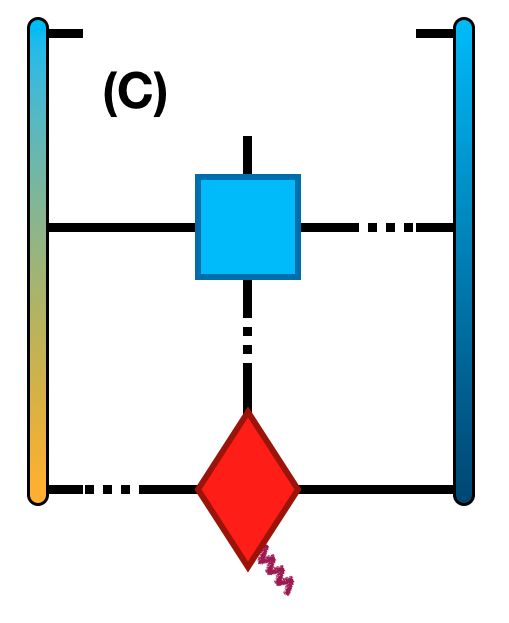}
\end{center}
\caption{Contraction order used for the single-site representation.  The left environment is contracted onto the MPO.  The right environment is contracted onto the bundled orthogonality center.  The last step is to take the two tensors and contract common indices.  In each step, tensors are contracted pairwise with the left and right given in the figure.  This order automatically places the indices in the same order as the input orthogonality center and avoids an extra permutation \cite{dmrjulia1}. The tensor notation here is given in Eq.~\eqref{Fig3tensdefs}.\label{contractionorder}}
\end{figure}

Implementing the block Lanczos routine into a tensor network is discussed here. Diagrammatically, Eq.~\eqref{eq:BL-rec} is represented in Fig.~\ref{blocklanczosfig} for the single-site case where the local Hilbert space size is of dimension $dm^2$ for physical index size $d$ and link index size of $m$ in the Lanczos procedure.  The additional index that records the current excitation is shown as a wiggly line on the diagrams. 

One item should be checked before beginning the above procedure.  The first step is to check that the MPS can solve for the requested number of excitations.  If, for example, the MPS has dimensions (2,2,4) and the number of excitations requested is greater than 16 (product of all dimensions), then the total Hilbert space on this site is not sufficient to generate those excitations.  If this is the case, the MPS is truncated on the excitation index to the maximal number.  

To normalize the bundled MPS tensor, an SVD can be used, grouping the excitation index on the right and the others on the left, but we discard the $\mathbf{D}$ matrix and contract only $\mathbf{U}$ and $\mathbf{V}^\dagger$ (Fig.~\ref{blocklanczosfig}a).  The input to the SVD is always normalized as in any Lanczos algorithm. This is done at each step in the recursion relation here to ensure that the matrices are composed of a fully orthonormalized set of vectors.  This also has practical use in real algorithms. If a non-normalized wavefunction is input to a DMRG algorithm ({\it e.g.}, starting from a purely random MPS or optimizing over $\hat c^\dagger_i|\psi\rangle$ without normalizing first for convenience) then this step will automatically create an appropriate tensor even if only some of the tensor was non-zero at the start.  This procedure of normalization has an advantage over dividing by a scalar in that a matrix composed of fully orthonormal rows and columns is produced.

We now explain the step in the diagrams. The center of orthogonality in the bundled MPS tensor is decomposed such that the three traditional indices of the MPS tensor (a physical index and two link indices \cite{bakerCJP21,*baker2019m}) are separated from the excitation index in the first step A.  The second step B is to contract the two unitary tensors together, forming the normalized input.  

The next three steps C--E are the diagram representations of the first three terms on the right hand side of Eq.~\eqref{eq:BL-rec}.  Step C is the MPO applied onto the MPS as would be traditionally expected in a single-site algorithm. The contractions necessary are shown in Fig.~\ref{contractionorder}. Step D applies the tensor representing $\mathbf{A}_n$ onto the excitation index of the bundle.  One can recognize the contraction of the diagram in step C to form the tensor that is applied on the excitation index.

The result of the diagrams in Fig.~\ref{blocklanczosfig} contains both $\Psi_{n+1}$ and $\mathbf{B}_{n+1}$.  The remaining task is to separate the two terms from each other so that the information can be used at the next iteration.  Note in diagram E where the component tensors for $\mathbf{B}$ were obtained.  These tensors are stored from the previous step of the recursion according to Eqs.~\eqref{eq:defB} and \eqref{nextPsi}.

The diagrams have a representation in a tensor format as
\begin{align}
\mathrm{Fig.~\ref{blocklanczosfig}a)}\quad&W^{\sigma_i\xi}_{a_{i-1}a_i}\overset{\mathrm{SVD}}=U^{\sigma_ip}_{a_{i-1}a_i}D_{pq}V_{q\xi}
\\
\mathrm{Fig.~\ref{blocklanczosfig}b)}\quad&A^{\sigma_i\xi}_{a_{i-1}a_i}=U^{\sigma_ip}_{a_{i-1}a_i}V_{p\xi}
\\
\mathrm{Fig.~\ref{blocklanczosfig}c)}\quad& (\psi_H)^{\sigma_i'\xi}_{a_{i-1}'a_i'}=(L^E)^{a'_{i-1}w_{i-1}}_{a_{i-1}}H_{w_{i-1}w_i}^{\sigma_i'\sigma_i}\label{Fig3tensdefs}\\
&\quad\quad\quad\quad\quad\quad\times A^{\sigma_i\xi}_{a_{i-1}a_i}(R^E)^{a'_iw_i}_{a_i}\nonumber
\\
\mathrm{Fig.~\ref{blocklanczosfig}d)}\quad& (\psi_A)^{\sigma_i\xi}_{a_{i-1}a_i}=A^{\sigma_ip}_{a_{i-1}a_i}A^{\sigma_i'p}_{a_{i-1}'a_i'}(\psi_H)^{\sigma_i'\xi}_{a_{i-1}'a_i'}\label{follow1}
\\
\mathrm{Fig.~\ref{blocklanczosfig}e)}\quad& (\psi_B)^{\sigma_i\xi}_{a_{i-1}a_i}=\bar A^{\sigma_i\eta}_{a_{i-1}a_i}\bar V_p^\eta\bar D_{pq}\bar V_{q\xi}\label{follow2}
\end{align}
where the $\times$ symbol merely denotes the continuation of the line. Note that the primed indices on $\psi_H$ in Eq.~\eqref{Fig3tensdefs} can be trivially renamed without the primes to match the following two tensors, Eqs.~(\ref{follow1}-\ref{follow2}). Tensors denoted with a bar are the tensors found at the previous iteration of the algorithm. The tensors $\psi_H$ correspond to the tensor generalization of the Hamiltonian acting on the input wavefunction, $\psi_A$ is the tensor generalization of $\mathbf{A}$ onto $\psi$, and $\psi_B$ is the tensor generalization of $\mathbf{B}$ onto $\psi$. Equal signs are denoted with SVD when an SVD is used, otherwise, the operation is a contraction.

Having constructed the block matrices, the super-matrix $\check M$ is formed as in Eq.~\eqref{blockM} and decomposed according to
\begin{equation}
\check M = \check U\check \Lambda\check U^\dagger.
\end{equation}
The diagonalization of this matrix yields a unitary $\check U$ that can then be applied onto the excitation index.  In our implementation, all states $\psi_i$ generated by the recursion equations are bundled into one tensor according to Sec.~\ref{bundleMPS}.  Then, $\check U$ is applied onto the excitation index of a bundled MPS tensor that has all excitations from previous steps. The number of excitations requested is then kept by truncating the excitation index.

This procedure is efficient for large models; however, one can imagine many possible permutations on this algorithm.

\subsubsection{Quantum numbers}

Quantum number symmetries are a useful concept in tensor networks and can drastically reduce the amount of computational time to perform operations.  In essence, each index of the tensors in the network can be labeled by a quantum number \cite{hauschild2018efficient,orus2019tensor}. The linear algebra operations are then decomposed into a series of smaller tensors, one for each quantum number sector.  This procedure effectively generates a block sparse representation of the tensors and can yield major savings in terms of efficiency of algorithms. For example, a contraction algorithm is divided into a set of smaller contractions, one for each quantum number value on the index being contracted. There is no explicit change to the steps in Fig.~\ref{blocklanczosfig} brought about by the the quantum number symmetries; however, there are some issues that require discussion.

If quantum number symmetries are specified for the excitation index, then the lowest energies in each sector will be found. Only excitations on the excitation index will be found in the normalization step, so these must be specified before performing any computations. Note that the lowest energies may not correspond to the quantum numbers of the excitation index.  If a fixed set of quantum numbers should be obtained after diagonalizing the block-tridiagonal Hamiltonian from Eq.~\eqref{blockM}, then the list of lowest energies must be identified by quantum number.  Then, only those lowest energies in the appropriate quantum number sector should be retained.

In the implementation here, a list of quantum numbers is simply passed to the block Lanczos routine and used to keep and sort the resulting excitations.  Alternatively, if no list is provided, then the lowest energies based on the input quantum numbers are kept.

\subsection{Two-site}

The standard version of DMRG is to use a reduced site representation of two sites.  Once the block Lanczos procedure for two sites is completed, the orthogonality center must be moved to the next site according to the diagrams

\includegraphics[width=\columnwidth]{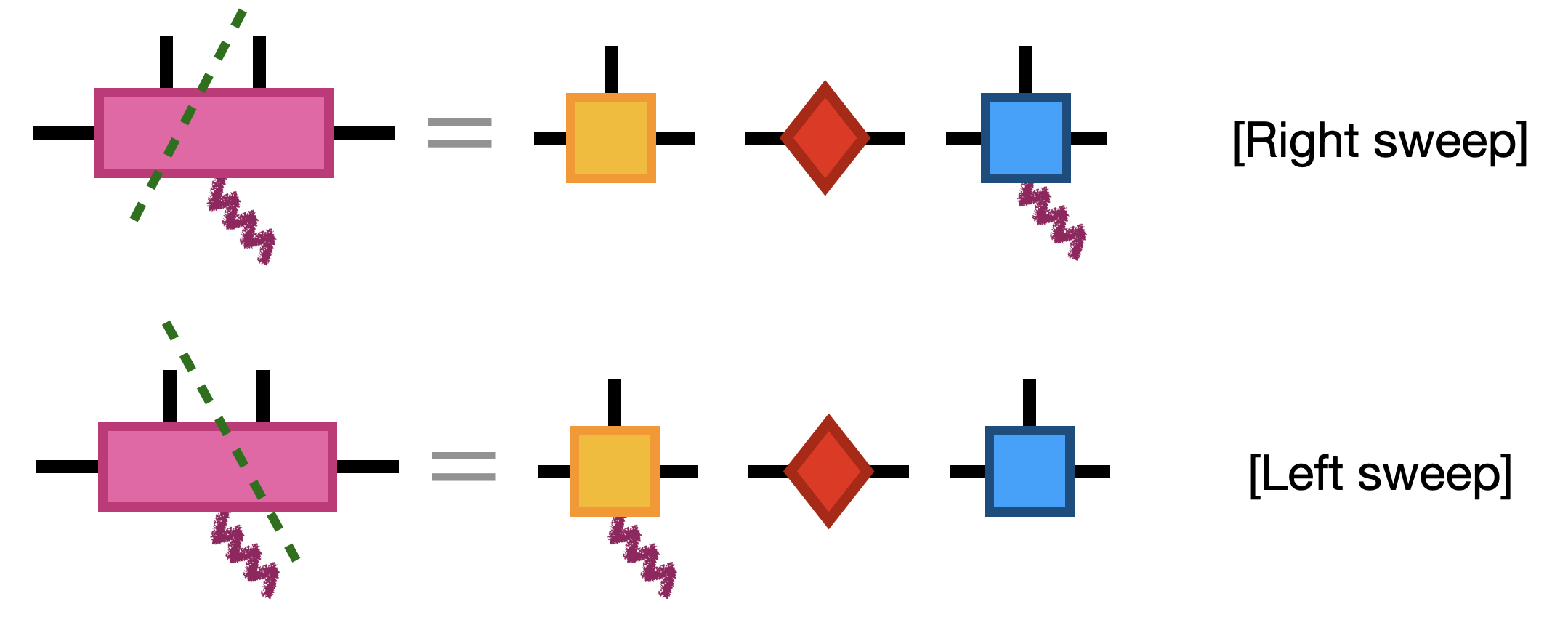}
where it should be noted that the starting tensor is the result of the block Lanczos procedure and the excitation index must be permuted appropriately for the SVD. This constitutes the only two replacements (Lanczos to block Lanczos and changing the movement operations) in the original DMRG algorithm that must be considered for the multi-targeted version presented here.

In a tensor format, the steps would be
\begin{equation}
A^{\sigma_{i-1}\sigma_i\xi}_{a_{i-1}a_i}\overset{\mathrm{SVD}}=U_{a_{i-1}p}^{\sigma_{i-1}} D_{pq}V^{\sigma_i\xi}_{qa_i}\quad\mathrm{[Right\;sweep]}
\end{equation}
and
\begin{equation}
A^{\sigma_{i-1}\sigma_i\xi}_{a_{i-1}a_i}\overset{\mathrm{SVD}}=U^{\sigma_{i-1}\xi}_{a_{i-1}p} D_{pq}V_{qa_i}^{\sigma_i}\quad\mathrm{[Left\;sweep]}.
\end{equation}

The cost of performing this contraction is $\mathcal{O}(d^2m^3g)$ per site (quadratically in the physical index size $d$ where $d^{N_s}=K$, cubically in the bond dimension $m$, and linearly in the number of excitations $g$).  Two-site methods are known to be stable and produce high quality ground states with no bias towards the direction of the sweep.  However, the introduction of the excitation index means that this algorithm will be more expensive. It should be noted that, even though it is more expensive, the local Hilbert space size is larger for the two-site method, so it is less prone to effects from truncation.  The natural question to ask is whether a single-site version can be introduced.  The answer is yes, and it turns out that it possesses almost no deficiency in comparison with the two-site implementation. So, use of the two-site algorithm is not necessarily required for this multi-targeted approach.

\subsection{Single-site}

Using more than one site to write the renormalized system can make the algorithm more expensive.  A single-site version is built here and shown to not have many of the deficiencies encountered for the single-site versions usually \cite{mcculloch2007density}.

The current best known algorithm for a single-site is the strictly single-site (3S) algorithm that has become standard in DMRG \cite{hubig2015strictly}.  This algorithm is reviewed in Appendix~\ref{backgroundinfo}, but the basic strategy still applies for the bundled MPS because the Hamiltonian is applied on each state in the bundle equally.

The statement made in Ref.~\onlinecite{hubig2015strictly} is to include a perturbation on the wavefunction that acts as the first order term in a Lanczos algorithm.  Essentially, this is $H|\psi\rangle$.  So, evolving the sites in the MPS according to $(H-E)|\psi\rangle$ will drive the wavefunction into the ground state. The perturbation can be modified with a noise term $\gamma$ as $|\psi\rangle\oplus\gamma H|\psi\rangle$.

For the multi-targeted algorithm, expressing the single-site operators in the left or right basis for the right or left sweep appears as

\includegraphics[width=\columnwidth]{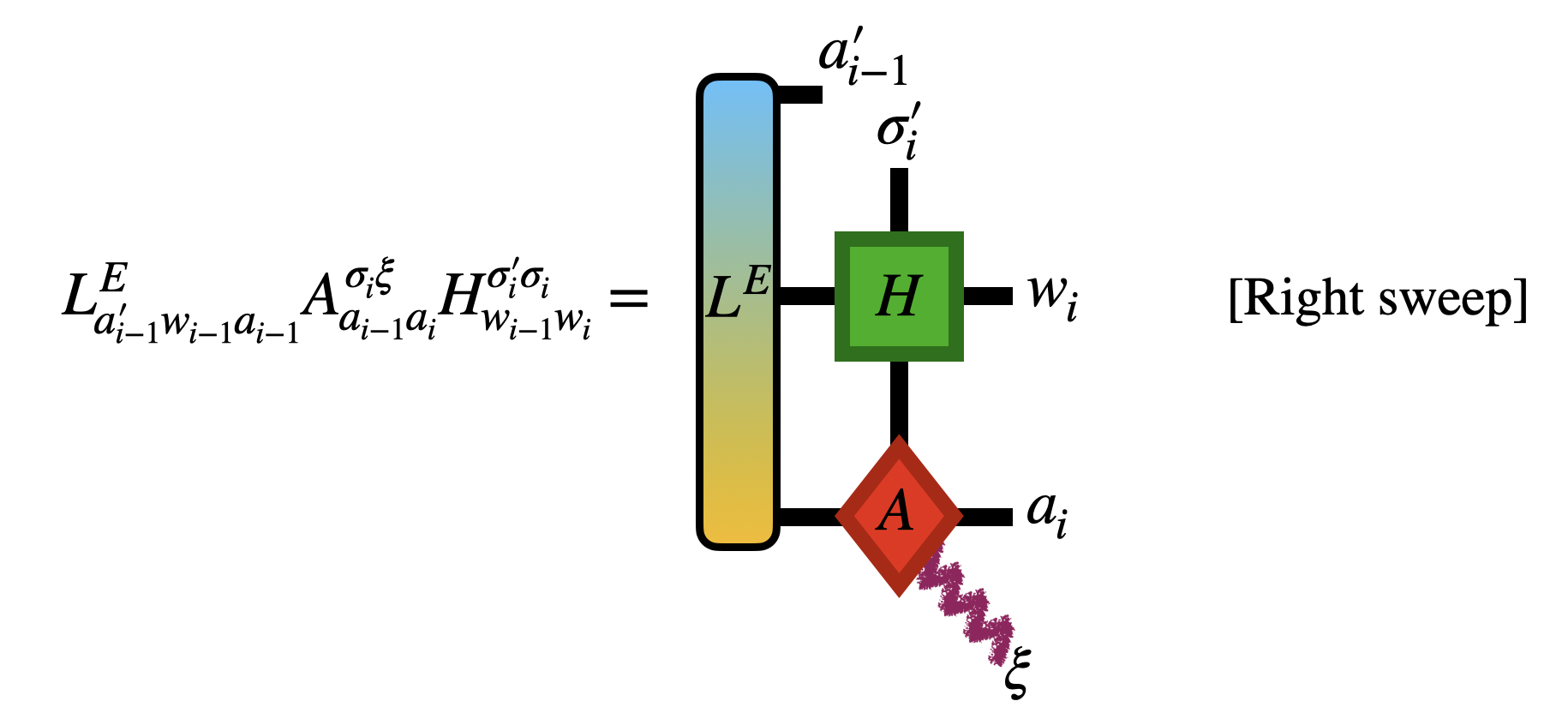}
\includegraphics[width=\columnwidth]{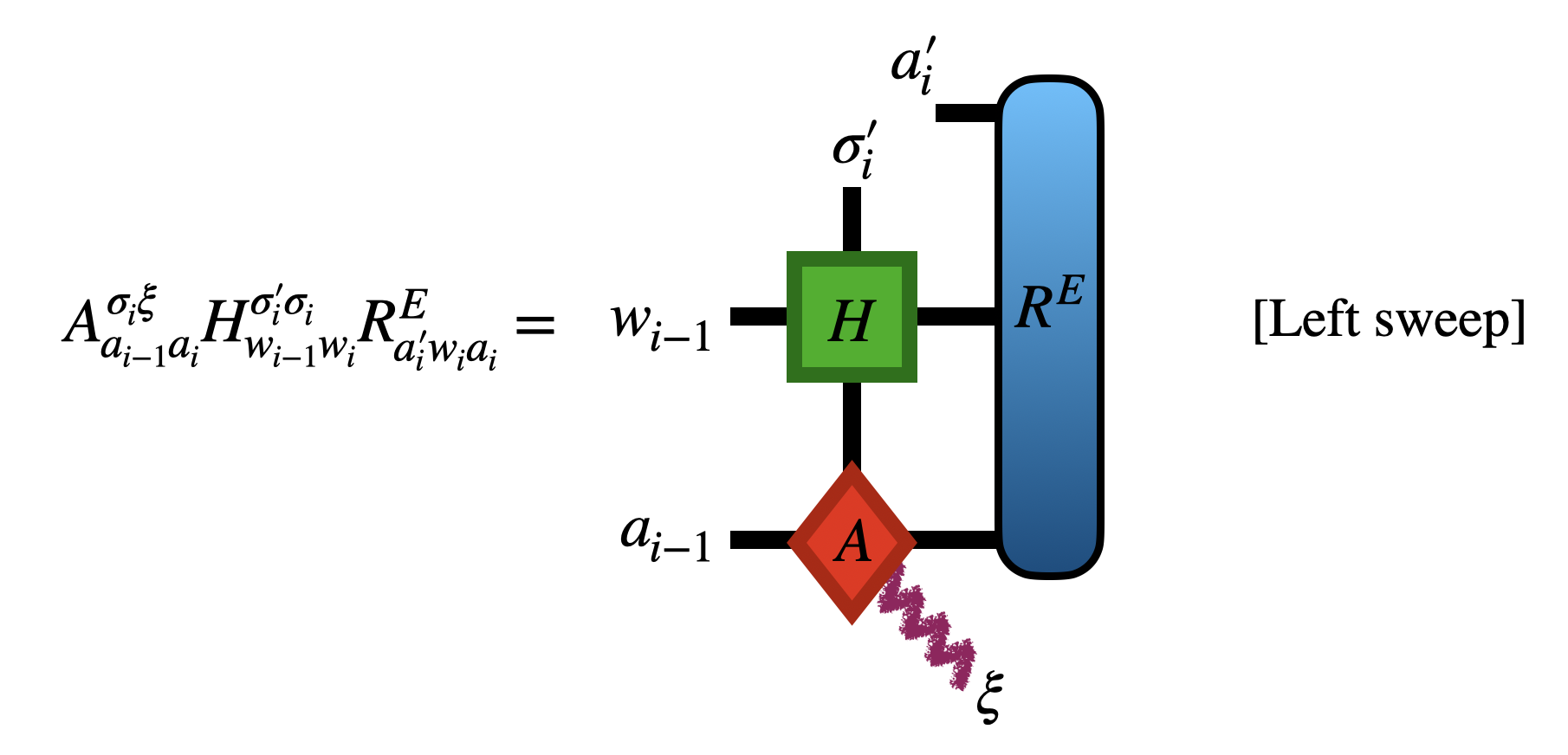}
where repeated indices are summed over and the resulting tensors are
\begin{align}
L^E_{a'_{i-1}w_{i-1}a_{i-1}}A^{\sigma_i\xi}_{a_{i-1}a_i}H^{\sigma_i'\sigma_i}_{w_{i-1}w_i}=X_{a'_{i-1}(a_iw_i)}^{\sigma_i'\xi}
\\
A^{\sigma_i\xi}_{a_{i-1}a_i}H^{\sigma_i'\sigma_i}_{w_{i-1}w_i}R^E_{a_i'w_ia_i}=Y_{(a_{i-1}w_{i-1})a'_i}^{\sigma_i'\xi}.
\end{align}
Two indices in parentheses, $(a_iw_i)$ for the first line and $(a_{i-1}w_{i-1})$ for the second, indicate that those indices were reshaped together and can be relabeled as a single index.

In the algebraic notation used here, the direct sum operator would appear for the bundled version as 
\begin{equation}
(A')^{\sigma_i\xi}_{a_{i-1}a_i}=A^{\sigma_i\xi}_{a_{i-1}\tilde a_i}\oplus X^{\sigma_i\xi}_{a_{i-1} \tilde a_i}
\end{equation}
which appears diagrammatically as

\includegraphics[width=\columnwidth]{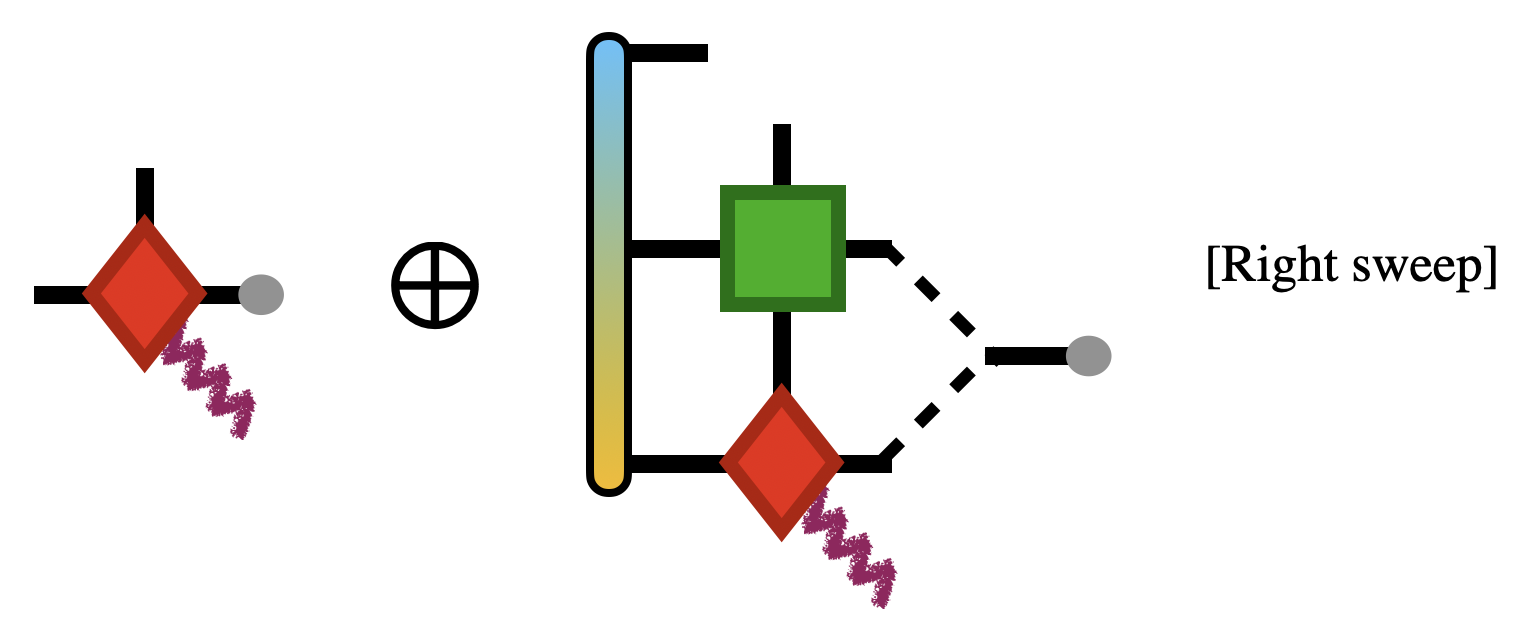}
similar to the ground state case.    For the other case when sweeping to the left,
\begin{equation}
(A')^{\sigma_i\xi}_{a_{i-1}a_i}=A^{\sigma_i\xi}_{\tilde a_{i-1}a_i}\oplus Y^{\sigma_i\xi}_{\tilde a_{i-1}a_i}
\end{equation}
which appears diagrammatically as

\includegraphics[width=\columnwidth]{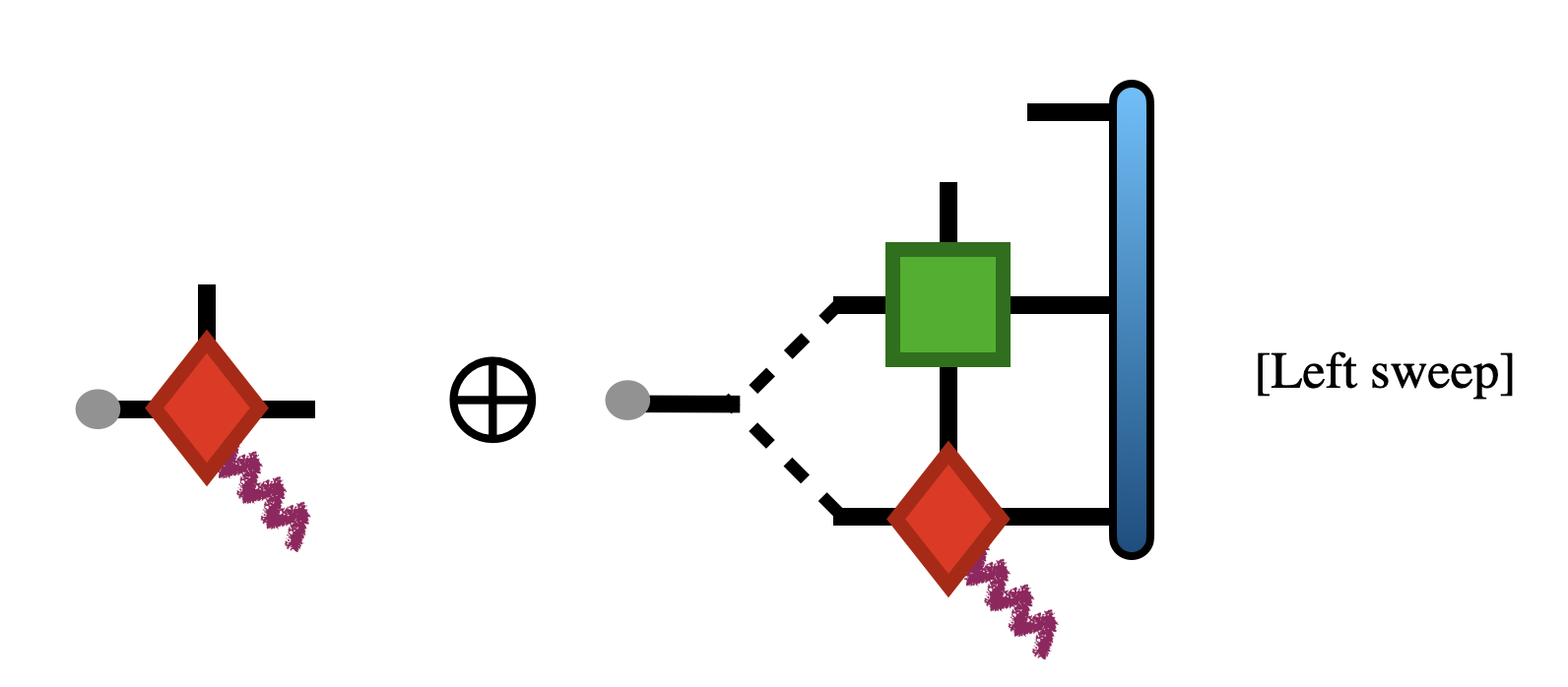}
where grey dots denote the indices that are being joined together.  The new $A'$ tensors can then be used in the algorithm in place of the original $A$ tensors, where the $A'$ tensors represent the original tensor joined with a representation of the Hamiltonian in either the left or the right environment ($X$ or $Y$ tensors).

The indices are truncated as is done in Ref.~\onlinecite{hubig2015strictly}, where the resulting $\mathbf{U}$ or $\mathbf{V}^\dagger$ tensor is truncated along the first or third index to match the size of the complimentary tensor. The cost of performing these contractions is $\mathcal{O}(dm^3g)$ for each site, a reduction of magnitude $d$ from the two-site version. 

The movement of the orthogonality center is given diagrammatically as

\includegraphics[width=\columnwidth]{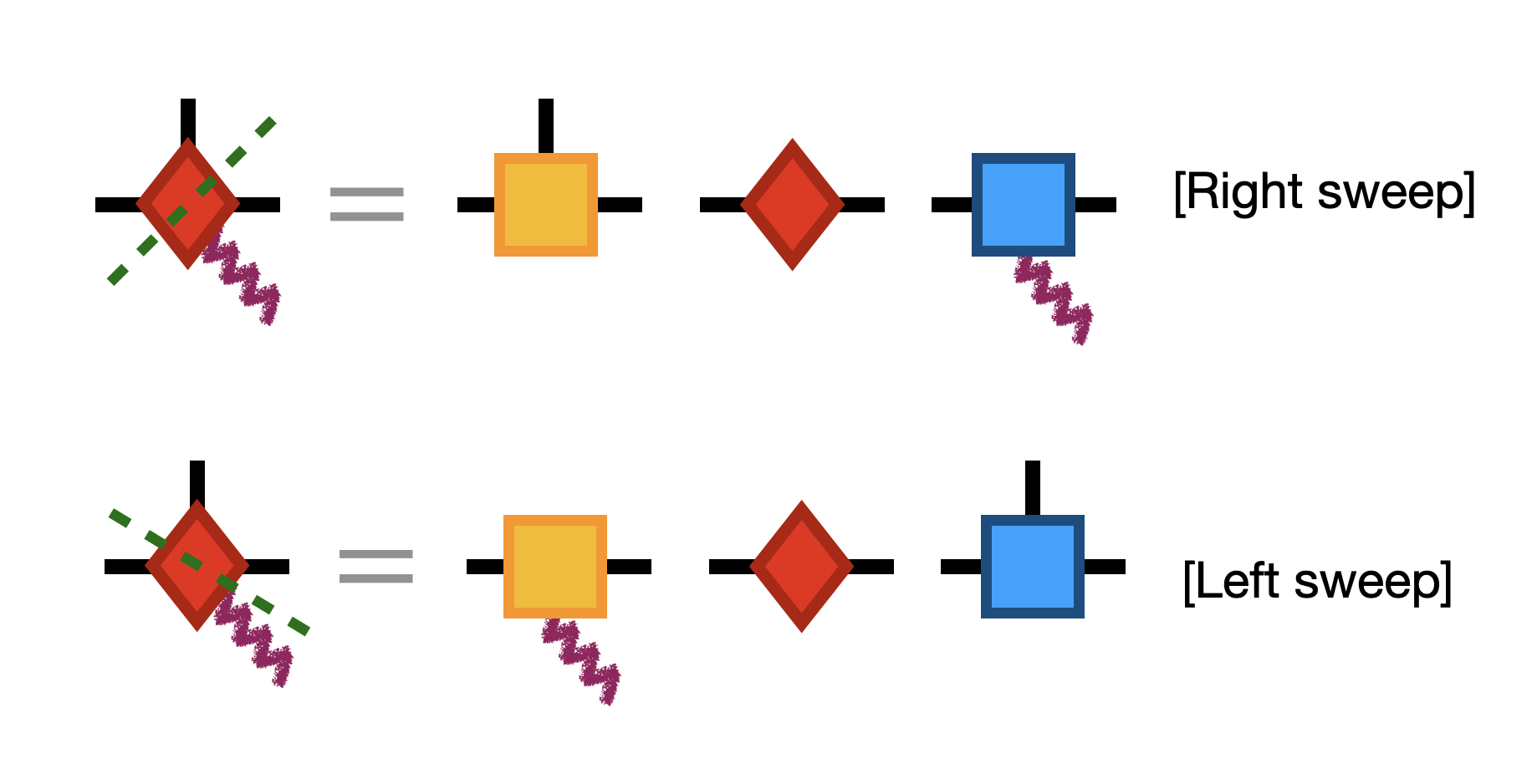}
where the excitation index is permuted appropriately to where the index must be moved. In a tensor notation, this is
\begin{equation}
A^{\sigma_i\xi}_{a_{i-1}a_i}\overset{\mathrm{SVD}}=U_{a_{i-1}p}^{\sigma_i} D_{pq}V_{qa_i}^\xi\quad\mathrm{[Right\;sweep]}
\end{equation}
and
\begin{equation}
A^{\sigma_i\xi}_{a_{i-1}a_i}\overset{\mathrm{SVD}}=U_{a_{i-1}p}^\xi D_{pq}V_{qa_i}^{\sigma_i}\quad\mathrm{[Left\;sweep]}
\end{equation}
In general, the additional index will always be attached to the orthogonality center throughout this work. 

\subsubsection{Noise term}

The algorithm presented so far assumes that a noise parameter $\gamma$ is always set to one.  This is not alway ideal for convergence and should be tuned.    The noise term from the 3S algorithm can be adapted here to improve convergence of the single-site.  This will provide a drastic improvement on the purely single-site bundle algorithm while keeping the cost savings in comparison with a two or more site algorithm.

In order to ensure that the noise terms are applied individually to each state, the excitation index is contracted onto a $g\times g$ diagonal matrix for $g$ excitations.  Each diagonal entry of the tensor contains the current noise term for that excitation.  Although, multiplying a single coefficient onto the tensor is possible without too much loss of performance in the algorithm.

\subsubsection{Entropy influenced by the noise term}

The addition of a noise term and appearance of many excitations can skew the entanglement measured on each bond.  Each excitation has its states stacked on top of each other, so all the lowest singular values are retained first without consideration for which excitation they belong to. 

The entanglement can therefore be strongly negative if the bundled MPS does not sufficiently converged and the noise term is still operating on the excitations.  The energies often converge before a non-negative entropy is found.

If the true entropy must be determined from the bundle, then the tensor representing the excitation of study should be copied from the bundle.  The SVD can be performed on this (regular) MPS tensor and the entropy can be determined.

The entropy we discuss here is computed over the full bond dimension during a move. If each excitation was selected individually, then the entropies of each excitation would be computed as normal with their normal behaviour.

\subsection{Summary of the multi-targeted algorithm}

To summarize the steps in the multi-targeted DMRG algorithm, the following list is provided.

\begin{enumerate}
\item Contract a tensor network (bundled MPS and MPO) down to a few-site system.
\item Perform a block or banded (or other matrix) Lanczos algorithm to update the tensors as in Eq.~\eqref{eq:BL-rec}. Only a requested number of excitations are kept.
\item The tensor is decomposed with the excitation index moved to the left or right depending on the sweep.
\item The center of orthogonality is moved to the next site according to the diagrams given here.
\end{enumerate}

\section{Demonstration on models}\label{models}

\begin{figure}[b]
\begin{center}
\includegraphics[width=\columnwidth]{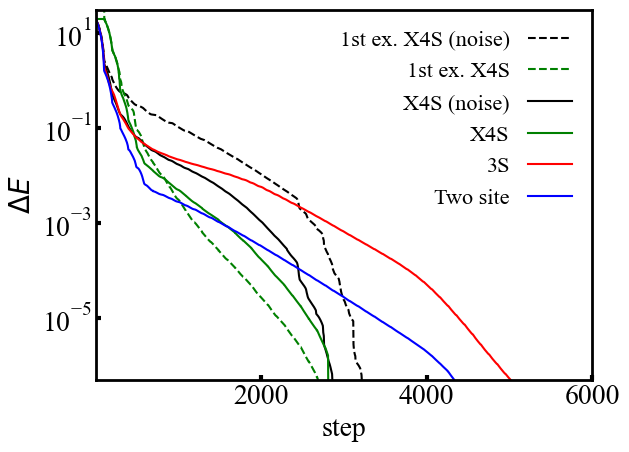}
\end{center}
\caption{Energy differences each time Lanczos is called in the traditional DMRG algorithm (two site), 3S algorithm, and the X4S introduced in the text. The model chosen is a 100-site spin-1/2 Heisenberg model with a cutoff (see Ref.~\onlinecite{dmrjulia1} for a definition) of $10^{-9}$, $m=200$, and only two excitations for the multi-targeted algorithm.  A noise term is optionally applied and while the convergence here is not as rapid beyond the first sweep or so, the noise term can be useful for the same reasons it was introduced in Ref.~\onlinecite{hubig2015strictly}.  \label{sitealgs}}
\end{figure}

Here the method is tested on several models to showcase some features that are present.

One point to ask that is not explicitly represented in the scaling of the method is how the bond dimension of the MPS scales as a function of the number of excitations requested.  Thus, in the worst case, one can expect that $m\propto g$.  However, this will be shown to be a non-standard case in the following.

\subsection{Ground state convergence}

Throughout these results, it can be seen that the method avoids many issues of previous methods for excited states, but it is useful to discuss the convergence of the ground state. The methods to be compared against are the two-site algorithm, the 3S algorithm, and the one introduced here.

It is natural to ask if this algorithm, which we will call an "excited state, striclty single-site" (X4S) algorithm here can improve on either the two-site algorithm or the 3S algorithm which was previously introduced into the literature.  To show this, the three algorithms are used to solve for the same system, a 100-site spin-1/2 Heisenberg model.  

The X4S algorithm does perform faster than the other algorithms that are compared against as demonstrated in Fig.~\ref{sitealgs}.  However, the accuracy of the method is somewhat limited if the same bond dimension is used because an additional excitation is present and the bond dimension cannot be screened successfully for the states only relevant to the ground state. It is possible to run the X4S, extract the ground state, and then continue running the two site or 3S algorithm to find a more accurate ground state, but we leave this to future users to determine if this is necessary for their problem. Some natural variation in the energy appears deep into the algorithm.  This is due to truncation of the bond dimension, especially at the edges, but these features disappear controllably with larger bond dimension.

\begin{figure}
\begin{center}
\includegraphics[width=\columnwidth]{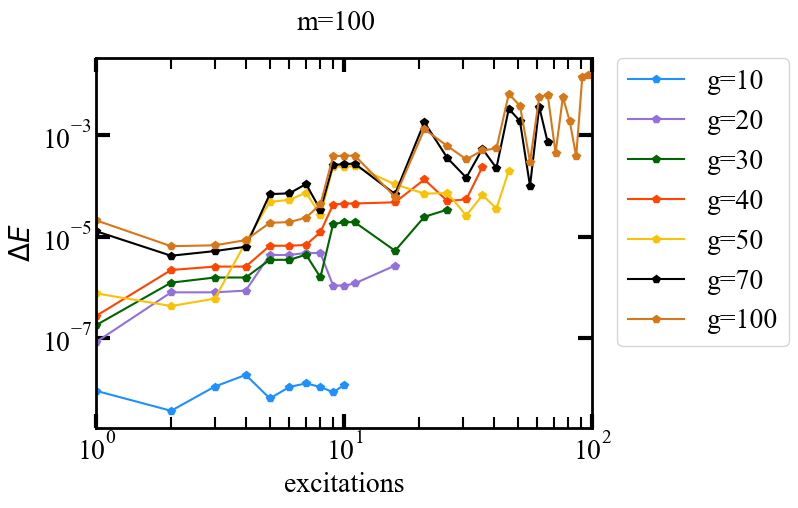}
\includegraphics[width=\columnwidth]{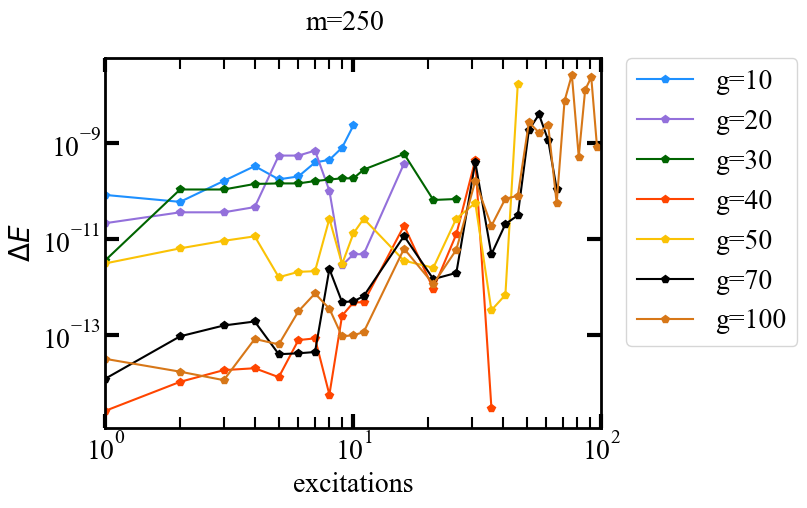}
\includegraphics[width=\columnwidth]{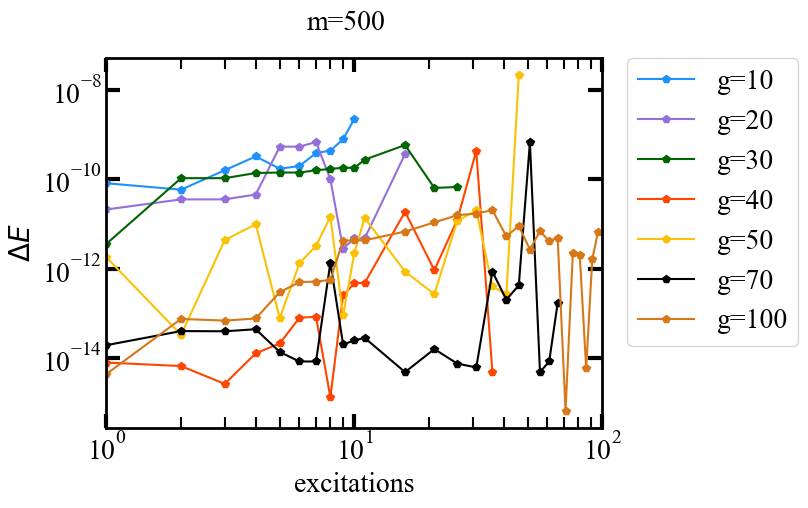}
\end{center}
\caption{Differences in energies for a requested number of excitations $g$ and for different maximum bond dimensions ($m=100, 250, 500$). The first ten excitations are shown and then every fifth excitation between 11 and 100.  The model is the 10-site spin-half Heisenberg model. Each set of excitations is solved to a given tolerance. With more sweeps, the bond dimension would decrease and errors would go to zero. \label{energydiffs}}
\end{figure}

\subsection{Excited state convergence}

With a decreasing bond dimension, the energy levels will become less accurate. To demonstrate this, Fig.~\ref{energydiffs} shows the energy differences for up to the first 100 excitations ($\sim10\%$ of the total Hilbert space size for a 10-site Heisenberg model) for three different bond dimensions ($m=100, 250, 500$, reported when the bundled MPS is gauged to the first site), which are set for each bond in the system.  The energies were solved to an individual accuracy of $10^{-8}$ as it is better to ensure each state converges rather than some grouped measure of the convergence.  There are several aspects to note.

First, the accuracy of the lowest energy level remains small no matter how many more of $g$ excitations are requested.  This is due to the largest singular values typically corresponding to the lowest energies of the system.  The higher lying excitations are only converged with increasing bond dimension, but simply increasing the bond dimension does not guarantee that the lowest excitations will be improved.  Rather, enough singular values pertaining to an excitation must not be truncated to get an accurate excited state energy.

With increasing bond dimension size, the excitations become more accurate, as can be seen in the decreasing energy differences of each graphs.  Also, higher lying excitations become more accurate because smaller singular values are kept in the SVD as the bond dimension increases. The differences in energies are far from monotonic.  This is because we have not screened the singular values for a particular excitation in this case, and such an operation would be difficult to perform efficiently.  The energies may be represented differently depending on the singular values available to them. If a higher bond dimension were kept, then more excitations could be retained.

Some variations in the accuracy of the eigenvalues appear as sharp jumps, albeit of a small magnitude. This is due to the natural variation in the singular values selected.  These jumps are repaired by keeping more singular values, although there is also the possibility to never truncate nearly degenerate singular values in the SVD.  This was implemented for the SVD in our computations and has a minimal effect, but may play an important role in future applications.

Note also that the ground state energy fluctuates slightly with an increasing number of excitations solved for.  Again, this is due to keeping the minimal singular values of the SVD.  Each of these graphs were found by keeping a noise term as explained earlier.  This aided convergence and allowed for a threshold of $10^{-12}$ to be reached in only a few sweeps (between 8 and 500 sweeps for a larger number of states in the bundle).  The code ran for no more than a few minutes depending on how many excitations were kept.  Memory was not an issue for a moderately sized systems, although the tensors were stored on the hard disk with little overhead for doing so.

The fact that the $g=10$ curve is similar for each figure is only based on the computation strategy used.  Since the excitations were only converged to a tolerance (and not a set number of runs), the computation will stop when the excitations are accurate.  This means that the extra states provided by the $m=500$ computation will converge in the same way as for the $m=250$ curve.  More accuracy can be obtained with more sweeps.  For the $m=100$ curves, the bond dimension is not large enough for the excitations and this generally produces poor convergence.  Again, increasing the bond dimension solves these problems.

It is remarkable that the $g=100$ excitations converge even for such a small bond dimension and that it can be solved quickly.  This greatly surpasses the penalty method introduced at the beginning of the paper.

\subsubsection{Bond dimension scaling}

As demonstrated in the last section, the bond dimension can be very small and still describe many excited states. This is due to the properties of the model in most cases and how many rows of a tensor reshaped into a matrix can store a set of indices for a given excitation.  It is conceivable that other models will not store the excitations so compactly, although our general experience with this algorithm is that this compactification is reliably found in more models.

However, the natural question to ask is how the bond dimension scales with increasing excitations.  Both empirically and logically the bond dimension appears to scale linearly with the number of excitations.  That would represent the worst case scenario, but it should be expected as demonstrated here that the bond dimension will often be much lower.

Note that if the algorithm were allowed to run much longer, then the optimal bond dimension for the excitations would be achieved. Further, all errors would go to zero as we will demonstrate on another model later. One strategy is to limit the bond dimension size to something reasonable until convergence and then check by raising it. The cutoff parameter will limit the growth of the total bond dimension size in many cases.

\subsubsection{Sparsity of link index basis}

Based on the findings of the previous section, it is a natural question to ask why the link indices are not growing exponentially with the number of excitations requested.  After all, this is the required bond dimension of the MPS for the single, exact wavefunction.  The answer is naturally that the low lying excitations share many basis states.  This section will estimate how many of these functions can be estimated in a spin model.

Taking results from exact diagonalization for a spin-half Heisenberg model with 30 sites, the results for the lowest energies in each of the lowest energies in the 16 distinct quantum number sectors can be obtained. Each sector differs by a total value of the sum of all of the $\hat S^z_\mathrm{tot.}=\sum_i\hat S^z_i$ values. While exact diagonalization captures all energies correctly, the same can be obtained by choosing an initial MPS in the correct quantum number sector and solving for the ground state with the symmetry enforced.  Doing so obtains excitations out to machine precision without the use of the multi-target algorithm here.  These excitations can be bundled together to obtain the exact bundle according to the algorithm presented in the earlier sections of this paper.

When the exact bundle is placed into the multi-DMRG algorithm, the SVD truncation removes some of the states that are necessary for describing high energy excitations in the bundle.  Depending on the details of truncation, only a certain number of excitations can be described by the basis. While the individual solutions in each of the quantum number sectors kept these, it must be minded that the weighting across different sectors can be different. Note that the exact bundle requires a bond dimension of approximately 700.

In this solution, a bond dimension of 200 allows for 8 excitations to be obtained.  Increasing the bond dimension to 1000 actually decreases the number of excitations found to 7 that can be solved for simultaneously (this is due to the uneven weighting of states when truncating with the SVD across all sectors). This implies that the low-lying excitations can be captured by a small bond dimension, but that higher excitations require larger singular values to be kept the SVD. If a worse starting state had been used ({i.e.}, only the lowest lying excitation), then only 7 excitations are found since it is difficult to obtain enough states on the non-excitation index to fully accommodate all solutions.

One way to visualize this is to consider the states necessary to describe the highest excitation where all spins are pointed up. The weight in the density matrix for the link index has a very small magnitude when weighted for the lowest excitation as this state contributes very little to the overall ground state. Hence, it is truncated.

In full, this is an inefficient way to use the algorithm since the excitations can be obtained by solving with the multi-target algorithm on one sector only. However, it does illustrate how close the relevant bond index states are to each other for this model. It should be noted that the method still performs extremely well for half of the possible excitations in this 30-site model. It should also be minded that the low-lying solutions within a given quantum number sector are easily found with a small bond dimension as demonstrated in other sections. 

\subsection{Degeneracy: Heisenberg models}

\begin{table}[t]
    \begin{tabular}{ |c|c|c|}
         \toprule
\hline
\multicolumn{3}{c}{\text{10-site, spin-half model}}\\
\hline
         
         \text{Deg.} & $E$ & $\Delta E$\\ 
         \hline\hline
1 & -4.2580352072798\hspace{0.1cm} & $1.0\times10^{-14}$\\
3 & -3.930673589(5)\hspace{0.3cm} & $4.4\times10^{-15}$\\
3 & -3.52704357161(5) & $1.0\times10^{-15}$\\
1 & -3.3961982689621\hspace{0.1cm} & $7.9\times10^{-15}$\\
3 & -3.1681508292(5)\hspace{0.2cm} & $1.8\times10^{-14}$\\
3 & -3.1505221075(2)\hspace{0.2cm} & $4.1\times10^{-15}$\\
1 & -3.0215944553979\hspace{0.15cm} & $2.6\times10^{-15}$\\
5 & -2.951230033(2)\hspace{0.4cm} & $8.0\times10^{-12}$\\
3 & -2.8874984638(9)\hspace{0.25cm} & $5.7\times10^{-14}$\\
3 & -2.8008740060(4)\hspace{0.25cm} & $6.8\times10^{-14}$\\
1 & -2.7298512548834\hspace{0.2cm} & $3.9\times10^{-15}$\\
3 & -2.709124428(7)\hspace{0.45cm} & $6.5\times10^{-14}$\\
5 & -2.61349518(1)\hspace{0.6cm} & $2.3\times10^{-10}$\\
1 & -2.5858046700255\hspace{0.2cm} & $4.4\times10^{-16}$\\
1 & -2.5447364827264\hspace{0.2cm} & $3.5\times10^{-15}$\\
3 & -2.527561848(0)\hspace{0.45cm} & $2.9\times10^{-13}$\\
3 & -2.41813085203(8)\hspace{0.15cm} & $2.3\times10^{-14}$\\
3 & -2.3538671946(5)\hspace{0.35cm} & $2.6\times10^{-13}$\\
5 & -2.34910996(8)\hspace{0.65cm} & $1.0\times10^{-9\;\,}$\\
         \hline
     \bottomrule \end{tabular}
\caption{The lowest 51 energies for the Heisenberg model with 10 sites and lowest differences with the exact values. Shown is the degeneracy (Deg.), and the number of digits consistent between the eigenvalues as determined from the solutions in DMRG (uncertain digits in parenthesis).  The bond dimension for this computation was $m=256$ with a cutoff \cite{dmrjulia1} of $10^{-9}$.\label{smallHeisenbergenergies}}
\end{table}

It is well known that the degeneracies of the solutions follow a specific pattern.  Namely, they are all odd numbers for the spin-$1/2$ chain \cite{haldane1983continuum,haldane1983nonlinear,lieb1989two,kramers1930theorie}. To demonstrate how reliable the degeneracies are recovered, the energies for each of 51 excitations are shown in Table~\ref{smallHeisenbergenergies}.  The energies differences with exact diagonalization show a small difference is found for a moderate bond dimension of $m=256$ on a 10-site Heisenberg model.

The differences in the digit precision of the values stated is greatly affected by the truncation of the bond dimension.  This is because the truncation of the bond dimension does not necessarily truncate all values of the index with any attention paid to the excitation.  Nevertheless, the number of digits in agreement for a given set of degenerate excitations are very high.  The digits in parentheses in Table~\ref{smallHeisenbergenergies} show which of the digits are uncertain for each of the degenerate states just by analysis of the multi-targeted DMRG solution.

Note that the difference with the exact value increases with degeneracy and as the excitation number goes up.  This can be attributed to the retention of the lowest eigenvalues in this case.  Since all eigenvalues are ordered over all excitations, the largest singular values are for the lowest energy levels.  This also means that solving for more excitations will increase the innacuracy in the low lying excitations in a straightforward application of this method.

The quality of the solutions are retained even for large lattices.  Table~\ref{bigHeisenbergenergies} shows the degeneracies and energies for the 100-site spin-half Heisenberg model.

\begin{table}
    \begin{tabular}{ |c|c|}
         \toprule 
\hline
\multicolumn{2}{c}{\text{100-site, spin-half model}}\\
\hline
         
         \text{Deg.} & $E$\\ 
         \hline\hline
1 & -44.12773930138912 \\
3 & -44.0872985(1)\hspace{0.7cm} \\
3 & -44.0395829(5)\hspace{0.7cm} \\
1 & -44.03100391578857 \\
3 & -43.9932390(4)\hspace{0.7cm} \\
3 & -43.9915532(4)\hspace{0.7cm} \\
1 & \hspace{0.2cm}-43.982462420453516 \\
5 & -43.9608810(7)\hspace{0.7cm} \\
         \hline
     \bottomrule \end{tabular}
\caption{The lowest 20 energies for the Heisenberg model with 100 sites. Shown is the degeneracy (Deg.), and the number of digits consistent within in each multiplet ({\it i.e.}, the number of digits in common between each eigenvalue with matching energy levels).  The bond dimension for this computation was $m=500$ with a cutoff \cite{dmrjulia1} of $10^{-9}$. More eigenvalues can be solved to similar accuracy, but the solution is more expensive. The presence of a noise parameter aided convergence far more rapidly than the noiseless case.
\label{bigHeisenbergenergies}}
\end{table}

\subsection{Low bond dimension: Fermionic models}

It is often the case that a quantum number can be identified as being conserved in a system.  A tensor network computation can accommodate a fixed number of particles for any Hamiltonian \cite{hauschild2018efficient,orus2019tensor}.  One of the most popular ways in which quantum numbers are implemented in MPS wavefunctions is to conserve fermion numbers.  This allows for the solution of a systems at fixed quantum number without varying a chemical potential.

The tight-binding ({\it i.e.}, non-interacting Hubbard) model can be written with constants such that it represents a discretized version of the continuum model \cite{bakerPRB15}.  In this case, the model will be solved without an interaction term to solve commonly known models, so a tight binding Hamiltonian is the result,
\begin{equation}
H=\sum_it\left(\hat c^\dagger_{i\sigma}\hat c_{i+1,\sigma}-\hat c_{i+1,\sigma}^\dagger\hat c_{i\sigma}\right)+\sum_i(\mu+v_i)\hat n_i
\end{equation}
where $\sigma=\{\uparrow,\downarrow\}$, $\hat n_{i,\sigma}=\hat c^\dagger_{i\sigma}\hat c_{i\sigma}$, $\hat n_i=\hat n_{i\uparrow}+\hat n_{i\downarrow}$, $t$ is a hopping parameter, and $\mu$ is a chemical potential. The external potential $v_i$ it chosen based on the system explored. To match the continuum results, we will use a grid spacing of $\Delta=1/N_s$, for $N_s$ grid points. This means that larger models will approach the thermodynamic limit.

There is a possibility to represent a continuum version of this model by setting \cite{wagnerPRB14}
\begin{align}
t=-1/(2\Delta^2)\quad\mathrm{and}\quad\mu=1/\Delta^2
\end{align}
which correspond to the second order finite different approximation of the continuum form of the second derivative operator in the kinetic energy, $\hat T=-\frac12\partial_x^2$ \cite{wagnerPRB14}.  The constant $\Delta$ is the grid spacing.  As the number of grid points is increased, the model will approach the continuum values of the full model.  The required bond dimension size will be small, but the ability for the multi-targeted DMRG method to resolve energy levels will continue.

In all simulations, a single electron is solved on $N>100$ sites, a moderate size.  There is no attempt to extrapolate the solution to the continuum limit, although there were no issues with increasing the number of lattice points in terms of the quality of the solutions found by the method.

For the large model, it was easiest to solve the lowest energy level in a regular (not bundled) MPS and then to explicitly bundle $g$ copies of the ground state together for the initial bundle for $g$ excitations.  This is necessary here because of the low bond dimension required for the solution with quantum number symmetries.  The normalization step automatically orthogonalizes all excitations in this case.  The bond dimension for these systems is extremely low ($m\approx5$ even with 4 excitations).  For many other cases of interest, where the bond dimension is larger, the normalization procedure generates an appropriate tensor to search for the excitation in another quantum number sector, provided that the quantum number labels allow for the correct block to appear.  We suspect that these low bond dimension results would be difficult to obtain with a post-processing technique for excitations, though we also have not seen an issue with larger bond dimensions for harder to solve systems that require larger bond dimensions.

\begin{table}
    \begin{tabular}{ |c|S[table-format=3.14, table-number-alignment=center]|c||c|}
         \toprule 
\hline
\multicolumn{4}{c}{\text{Particle in a box}}\\
\hline
         
         \text{$\nu$} & \text{$E_\nu$} & $\Delta E$ & \text{Theory}\\
         \hline\hline
1 & 0.980410765687 & $4.7\times10^{-12}$ & 1\\
2 & 3.920713083753 & $1.6\times10^{-12}$ & 4\\
3 & 8.818117899345 & $3.7\times10^{-12}$ & 9\\
4 & 15.667979727343 & $3.2\times10^{-12}$ & 16\\
5 & 24.463801058885 & $2.6\times10^{-12}$ & 25\\
6 & 35.197238524645 & $2.7\times10^{-12}$ & 36\\
7 & 47.858110809030 & $1.2\times10^{-12}$ & 49\\
8 & 62.434408307765 & $4.7\times10^{-12}$ & 64\\
9 & 78.912304519741 & $1.6\times10^{-12}$ & 81\\
10 & 97.276169162279 & $2.1\times10^{-12}$ & 100\\
         \hline
     \bottomrule \end{tabular}
\caption{
The first 10 energies and their difference with exact energies, $E_\nu$, are given in this table for a single particle in a box. Parameters are chosen such that ($\frac{\pi^2}{2L}=1$ and also that $\hbar=m_e=1$, for Planck's constant and the electron mass). The number of grid points $N_s=101$ are chosen so that the center point can handle a node more easily. The grid spacing is $\Delta=1/N_s$. Deviations from the theoretical values are on the order of $10^{-4}$ are due to the finite-difference approximation of the kinetic energy operator on a grid, not to the algorithm since it matches a full diagonalization. The values labeled by $\Delta E$ correspond to the value in the basis as found from the diagonalization of the single-particle Hamiltonian on the discretized grid (including some discretization error), and the theoretical value is reported in the continuum limit.\label{particleinaboxtable}}
\end{table}

\begin{figure}
\includegraphics[width=\columnwidth]{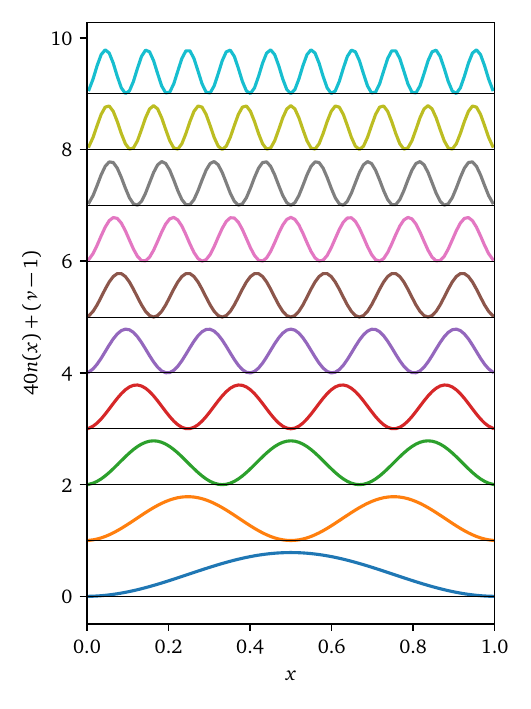}
\caption{Densities $n(x)=\langle \hat n_i\rangle$ (shifted by the value $2(\nu-1)$) for a particle in a box. The bundled MPS was solved with $m=100$ many body states kept and all excitation solved simultaneously.}
\label{particleinabox}
\end{figure}

Note that even for higher energy excitations that the boundary conditions behave smoothly at the edges.  This is true for the Heisenberg model presented earlier as well. This implies that the solution of the energy eigenstates is not suffering from numerical precision issues as more excitations are demanded.

\subsubsection{Particle in a box}

In the case of a particle in a box, $v_i=0$ everywhere.  The values for the eigenstates correspond nearly to the theoretical values $E_\nu=\nu^2\pi^2/(2L^2)$ in the continuum and should match within the tolerance of the chosen grid spacing, $\Delta$.  Both the energies and the one-body densities are shown in Fig.~\ref{particleinabox} and Table~\ref{particleinaboxtable}.

In both cases, the eigenvalues and eigenvectors match perfectly with exact diagonalization on the same lattice, even for the larger lattice sizes.

\subsubsection{Quantum harmonic oscillator}

\begin{figure}
\includegraphics[width=\columnwidth]{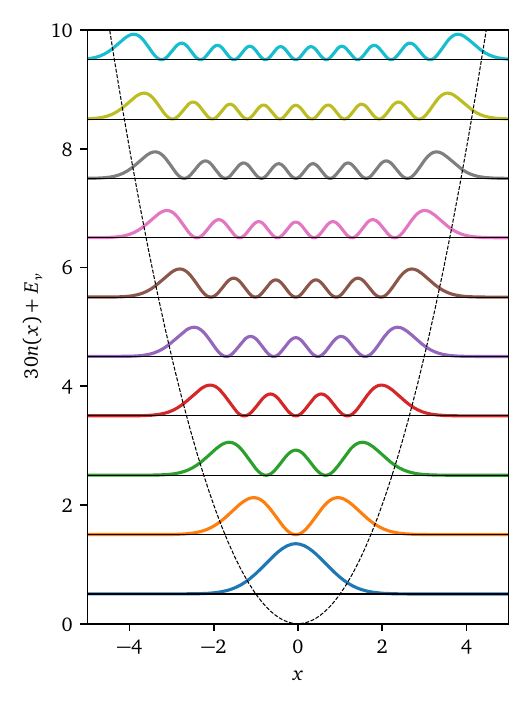}
\caption{Densities $n(x)=\langle \hat n_i\rangle$ (shifted by the exact energy value $E_\nu$) for the quantum harmonic oscillator. The bundled MPS was solved with $m=11$ many body states kept and all excitation solved simultaneously.}
\label{QHOdensities}
\end{figure}

The quantum harmonic oscillator is defined when a potential of the form
\begin{equation}
v_i=\frac12\omega^2\Big(\Delta (i-N_\mathrm{center})\Big)^2
\end{equation}
where the frequency $\omega$ is set to 1 throughout this demonstration and $N_\mathrm{center}=(N_s-1)/2$.  The values for the eigenstates correspond identically to the theoretical values $E_\nu=\omega(\nu+\frac12)$ up to a tolerance of the chosen grid spacing, $\Delta$.  Both the energies and the one-body densities are shown in Fig.~\ref{QHOdensities} and Table~\ref{QHOenergies}.

\begin{table}
    \begin{tabular}{ |c|S[table-format=2.17, table-number-alignment=center]|c||c|}
         \toprule 
\hline
\multicolumn{4}{c}{\text{Quantum harmonic oscillator}}\\
\hline
         
         \text{$\nu$} & \text{$E_\nu$} & $\Delta E$ & \text{Theory}\\
         \hline\hline
0 & 0.499921862787260 & $3.6\times10^{-14}$ & 0.5\\
1 & 1.499609265050815 & $3.9\times10^{-14}$ & 1.5\\
2 & 2.498983947278260 & $2.7\times10^{-14}$ & 2.5\\
3 & 3.498045762526104 & $9.3\times10^{-14}$ & 3.5\\
4 & 4.496794563620553 & $3.1\times10^{-14}$ & 4.5\\
5 & 5.495230203156738 & $3.9\times10^{-14}$ & 5.5\\
6 & 6.493352533498261 & $9.7\times10^{-14}$ & 6.5\\
7 & 7.491161406776690 & $3.3\times10^{-14}$ & 7.5\\
8 & 8.488656674890615 & $3.0\times10^{-14}$ & 8.5\\
9 & 9.485838189505555 & $1.4\times10^{-13}$ & 9.5\\
         \hline
     \bottomrule \end{tabular}
\caption{The first ten energy levels, $E_\nu$ of the quantum harmonic oscillator ($\omega=1$). The values were found on a 501 site lattice with a grid spacing of $\Delta=0.05$. Shown is the difference from the answer from exact diagonalization, $\Delta E$, and the number of digits consistent between the eigenvalues.  The bond dimension for this computation was $m=256$ with a cutoff of $10^{-9}$.
\label{QHOenergies}}
\end{table}

The method tends to perform well even for many excitations.  Out to the number of excitations shown, there is no discrepancy between the eigenvalues and eigenvectors from exact diagonalization.

\subsubsection{Discussion}

The models in this section are standard examples in quantum physics and provide a baseline for how the method performs with a single particle.  These examples do not contain an interaction, making them somewhat atypical example uses of tensor network methods, but they convey the abilities of these methods.

We note one feature of both Fig.~\ref{particleinabox} and Fig.~\ref{QHOdensities}. The results are relatively symmetric after a few sweeps. It has long-been known that the 3S method and other single-site methods can bias the results in one direction or another, but the X4S algorithm appears to correct these biases with subsequent sweeps.

In comparison with other techniques, namely Ref.~\onlinecite{chepiga2017excitation}, our method found a more consistent agreement with the eigenenergies for all eigenvalues. The memory cost savings of Ref.~\onlinecite{chepiga2017excitation}'s method did not necessarily make up for the loss of precision in our view. We also have not noticed any loss of degeneracies with this method even for very large systems \cite{di2021efficient}. One additional issue is that the edges of the system are well behaved here and lack oscillations. While the worst case scenario of linear scaling with the number of excitations for the bond dimension is given, it is shown here in practice that the optimal bond dimension is often far less than this cutoff \cite{dmrjulia1}.

While fermionic systems were studied here, large bosonic systems with applications to superconducting qubits can also be handled with this method \cite{di2021efficient}.

\paragraph*{Note added after submission} We were alerted to the prior publication of a similar method in Ref.~\onlinecite{dolgov2014computation} which is based on the LOBPCG method from Ref.~\onlinecite{knyazev2001toward}. The block Lanczos technique here is drastically simpler. We note that the algorithm here does not require the merging of blocks after solution by block Lanczos technique (as is stated in Algorithm 1 of Ref.~\onlinecite{dolgov2014computation}). Rather, the method here solves for excitations simultaneously. Tensors corresponding to individual eigenvalues do not necessarily need to be orthogonalized against each other (as just before Eq. (9) of Ref.~\onlinecite{dolgov2014computation}) but that we chose an equivalent SVD based approach here to compensate for the noise term. We also note that the single-site noise term to aid convergence is novel to this work. The method here does not precondition the eigenvalue problem, but this is often unnecessary for the update procedure of the matrix product state. The method in this paper has also been evaluated far beyond 100 excited states, the stated limit of Ref.~\onlinecite{dolgov2014computation}. It is not clear whether the other algorithm resolves degeneracies as well as the present algorithm.

\section{Conclusion}

A multi-targeted DMRG algorithm was described.  The main ingredients were to replace the Lanczos step in DMRG with a block Lanczos procedure and then to modify the movement operations.  The MPS ansatz was extended to the case of a bundled MPS that included an extra index to catalog excitations.

The method was able to resolve degeneracies without increasing the MPO's bond dimension, as is the case with the penalty method mentioned in the introduction.  The ability for the multi-targeted block Lanczos algorithm to discover excited states without too much computational overhead is a great advantage over the penalty terms that are the most common way to introduce excitations into the MPS at the current time.

The multi-targeted algorithm is able to resolve excitations simultaneously, does not increase the bond dimension of the MPO, and does not suffer from systematic issues with degenerate states, even in large systems. The method works on either small or large bond dimension problems, particularly those in superconducting circuits, and has has been reliable even in hard to solve models.

\section{Acknowledgements}

This research was enabled in part by support provided by Calcul Qu\'ebec (www.calculquebec.ca) and Compute Canada (www.computecanada.ca). 

This project was undertaken on the Viking Cluster, which is a high performance compute facility provided by the University of York. We are grateful for computational support from the University of York High Performance Computing service, Viking and the Research Computing team.

This work has been supported in part by the Natural Sciences and Engineering Research Council of Canada (NSERC) under grants RGPIN-2015-05598 and RGPIN-2020-05060.

T.E.B.~thanks funding provided by the postdoctoral fellowship from Institut quantique and Institut Transdisciplinaire d'Information Quantique (INTRIQ). This research was undertaken thanks in part to funding from the Canada First Research Excellence Fund (CFREF). 

T.E.B.~is grateful to the US-UK Fulbright Commission for financial support under the Fulbright U.S. Scholarship programme as hosted by the University of York.  This research was undertaken in part thanks to funding from the Bureau of Education and Cultural Affairs from the United States Department of State.

This research was undertaken, in part, thanks to funding from the Canada Research Chairs Program. The Chair position in the area of Quantum Computing for Modelling of Molecules and Materials is hosted by the Departments of Physics \& Astronomy and of Chemistry at the University of Victoria. T.E.B. is grateful for support from the University of Victoria's start-up grant from the Faculty of Science. 

\section{Author Contributions}

A.P.-F.~conceived of the general strategy using block Lanczos and made the first implementation of the two-site algorithm. T.E.B.~implemented all other elements and wrote the paper. D.S.~provided support and introduced the concepts of the block Lanczos algorithm to the other authors and provided reference data to check the algorithm.

\section{Code Availability}

Code used for this paper is derived entirely from the DMRjulia library for tensor network computations \cite{dmrjulia}.

\begin{appendix}

\section{Review of matrix product states and the density matrix renormalization group}\label{backgroundinfo}

The relationship between the MPS and the full wavefunction for a quantum system can be seen as a series of tensor decompositions on the full wavefunction \cite{dmrjulia1}. Recasting an $N$-site wavefunction as a tensor, $\psi_{\sigma_1\sigma_2\ldots\sigma_N}$, with physical degrees of freedom expressed by $\sigma_i$, then a series of QR decompositions can give the state \cite{bakerCJP21,*baker2019m}
\begin{equation}
|\psi\rangle=\sum_{\{a_i\},\{\sigma_i\}}A^{\sigma_1}_{a_1}A^{\sigma_2}_{a_1a_2}\ldots D^{\sigma_N}_{a_{N-1}}|\sigma_1\sigma_2\ldots\sigma_N\rangle
\end{equation}
where a series of reshaping operations group and separate the indices so that the QR decomposition on matrices can be used. Raising and lowering an index has no significance in this notation or in general for a tensor network \cite{bakerCJP21,*baker2019m}. An extra index $a$ is introduced and corresponds to the remaining states in a system \cite{bakerCJP21,*baker2019m}, as introduced during a tensor decomposition.  The final tensor is denoted as $D$ to signify the orthogonality center, which is the only tensor in the network that does not contract to the identity and contains the square root of the coefficients of the density matrix for whatever partition is selected in the system \cite{dmrjulia1}.  The tensors $A$ are left-normalized and when contracted on the dual left-normalized tensors, contract to the identity. The gauge of an MPS can be changed by a series of SVDs (or QR or LQ decompositions \cite{dmrjulia1}) which transforms some of the tensors to right-normalized tensors, $B$.  No matter how the MPS is gauged, a single orthogonality center is present.

The challenge for handling many excitations is to incorporate each as an additional index on the orthogonality center.

\subsection{Two-site density matrix renormalization group algorithm}

The DMRG algorithm operates on an MPS and has the following four basic steps in the algorithm.  These are repeated until convergence \cite{dmrjulia1}.

\begin{enumerate}
\item Contract a tensor network (MPS and MPO) down to a few-site system.
\item Perform a Lanczos algorithm to update the tensors
\begin{equation}\label{lanczos}
|\psi_{n+1}\rangle=H|\psi_n\rangle-\alpha_n|\psi_n\rangle-\beta_n|\psi_{n-1}\rangle
\end{equation}
with $\alpha_n=\langle\psi_n|H|\psi_n\rangle$ and $\beta_n^2=\langle\psi_{n-1}|\psi_{n-1}\rangle$.  The coefficients form a tridiagonal matrix which is the Hamiltonian in the Lanczos basis \cite{senechal2008introduction,baker2021lanczos}.
\item The tensor is decomposed with--for example--the SVD to break apart the tensors
\item The center of orthogonality, $\hat D$, is moved onto the next tensor.
\end{enumerate}
Having found the renormalized representation of the problem for the next subset of sites by updating the environment tensors for the reduced site representation, the steps can be repeated.  Once the full lattice is swept by the algorithm, one iteration is complete and the algorithm can be run until the ground state converges or some other criterion is met.  More details (and diagrams) are found in Ref.~\onlinecite{dmrjulia1}.

The Lanczos procedure in step 2 is typically only done for a few iterations ({\it i.e.}, 1).  This is because the environment contracted down to in the first step is typically not correct. Thus, a small update to the local tensors is useful in this case. The Lanczos step could be replaced by another method, but this is the generally accepted method to update the tensors in DMRG.

Originally, the algorithm was demonstrated for a renormalized problem of 2 sites.  However, there is one variation of this algorithm for a single-site that is worth mentioning here since it can drastically reduce the total computational time.  The time for traditional DMRG is $O(d^2m^3)$ per site with size $d$ physical index and a bond dimension of $m$ \cite{dmrjulia1}.

\subsection{Strictly single-site (3S) algorithm}\label{3Soverview}

It is possible to formulate DMRG such that it only uses one site.  The first attempt we will discuss here that is truly a single-site algorithm was from Ref.~\onlinecite{mcculloch2007density}.  There were some systematic issues with this method, so Ref.~\onlinecite{hubig2015strictly} introduced a correction that is widely used at the present time.  This modified single-site algorithm (called the strictly single-site algorithm or 3S) relies on an expansion of the tensor along the first Krylov vector of a Lanczos algorithm.

Steps 1 and 2 are kept the same in this DMRG algorithm from above.  Once the single-site is updated with Lanczos, another state, $(H-E)|\psi\rangle$ is added onto the single-site tensor (with eigen-energy $E$).  This is the first iteration of the Lanczos algorithm from Eq.~\eqref{lanczos}.  This requires that the tensor representing the site and a new tensor with the associated MPO contracted onto it are joined together using the $\oplus$ operation defined in the main text for the bundled MPS.

The original problem was represented in tensor notation, but the diagrammatic version will be shown here.  Because the MPO contracted onto the MPS tensor has extra indices, it is best to take the expansion term $H-E$ in the left and right environments depending on which direction the DMRG algorithm is sweeping in.  For the left sweep the operation is

\includegraphics[width=\columnwidth]{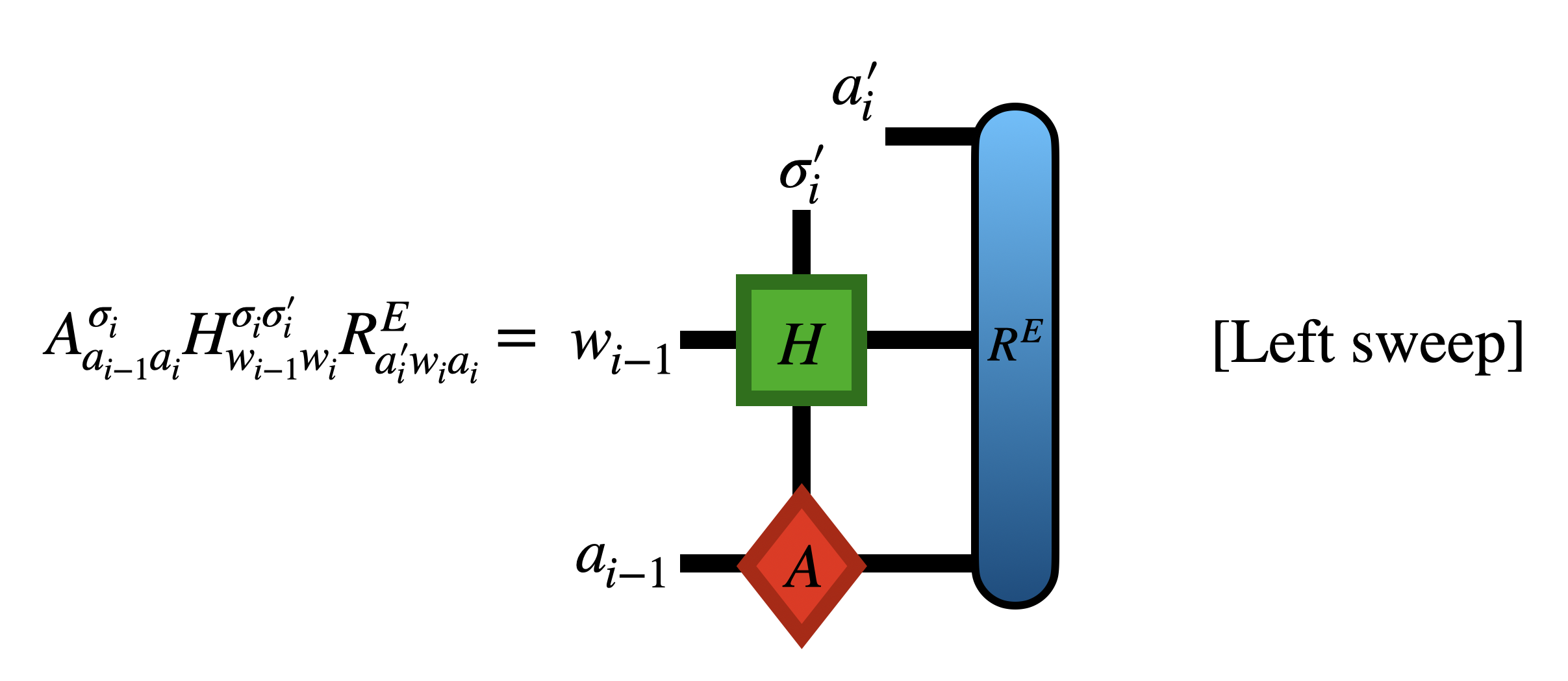}
and for the right sweep the operation is

\includegraphics[width=\columnwidth]{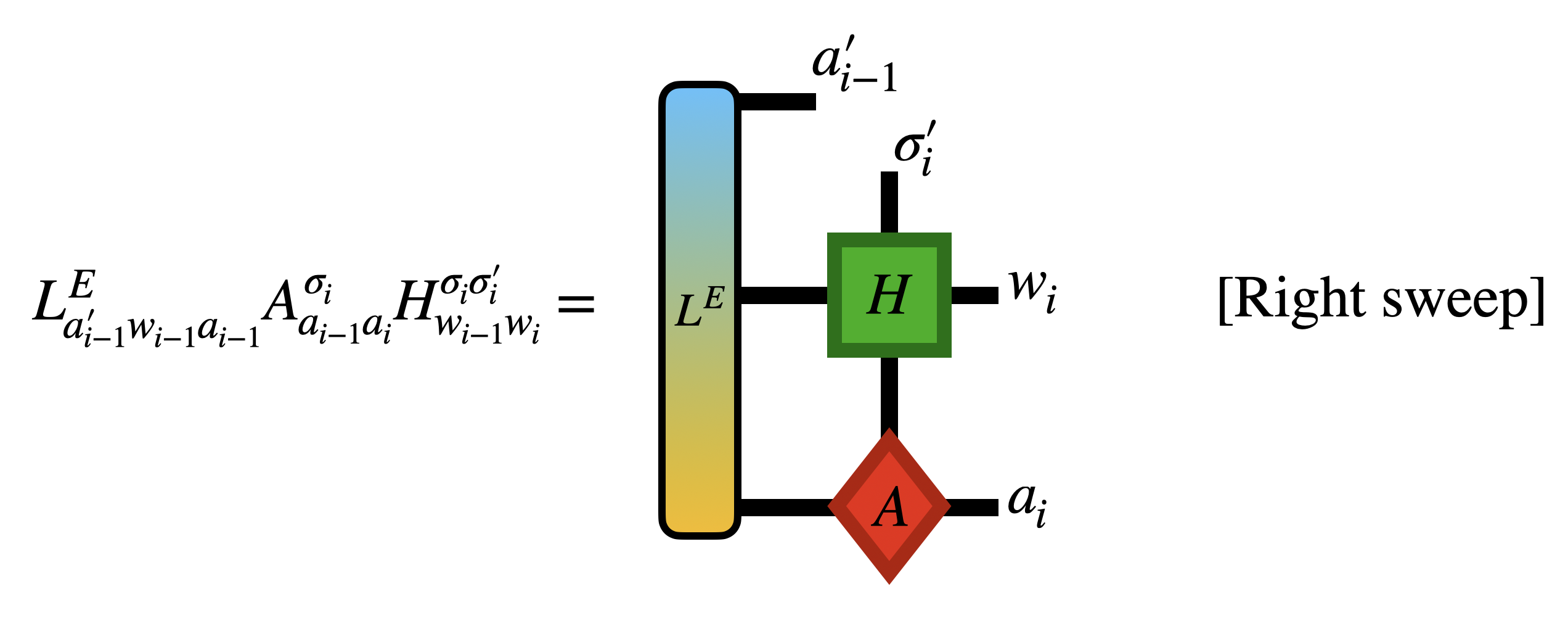}
and the MPO can be contracted inside of the environment ($L^E$ for the left sweep, $R^E$ for the right).  This forms the term $H-E$ in the appropriate basis.

The next step is to join the link indices of the MPS and MPO together to then join those indices to the original tensor according to the following diagrams

\includegraphics[width=\columnwidth]{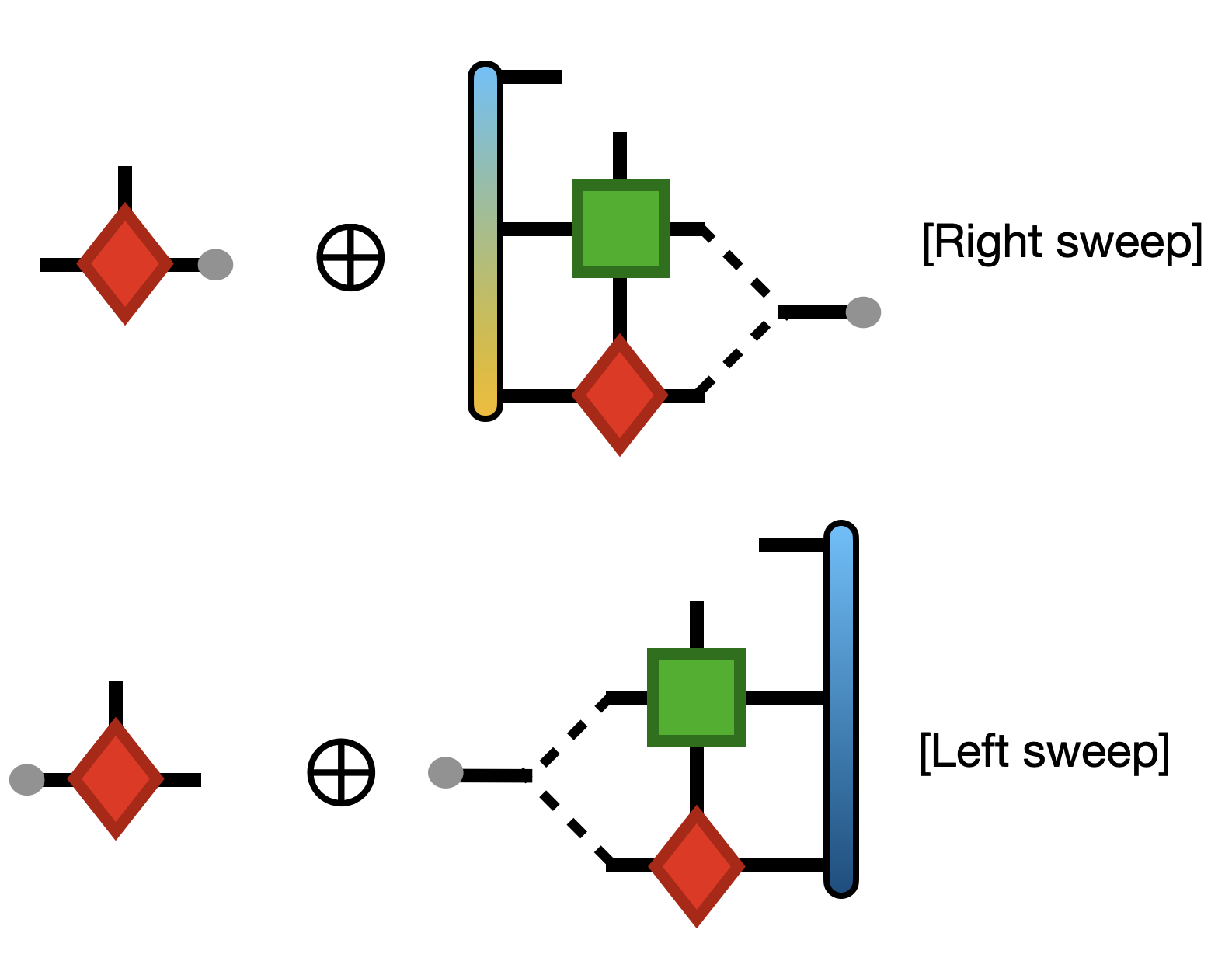}
where grey dots represent indices that are joined together.  Thus, the dotted indices are joined together into a single index, similar to a reshape but adding the dimensions together instead of maintaining the same number of elements as in a reshape \cite{bakerCJP21,*baker2019m}.

The final step is to move the orthogonality center according to the following diagrams

\includegraphics[width=\columnwidth]{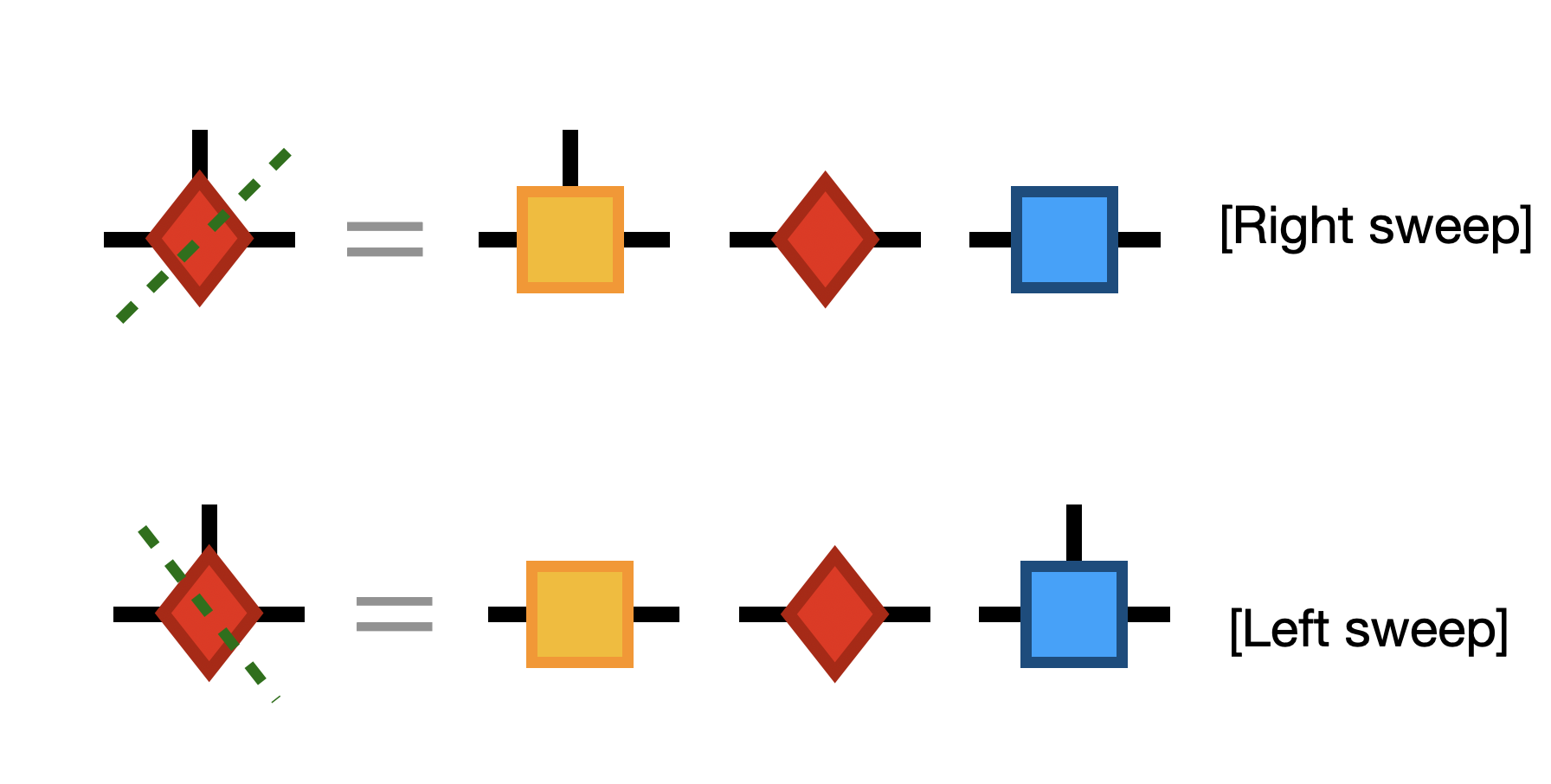}
where the two tensors on the right are contracted onto the next MPS tensor for the right sweep.  Similarly, the two tensors on the left of the decomposition are contracted onto the tensor on the left for the left sweep.

In the tensor form, the above equations are
\begin{equation}
A^{\sigma_i}_{a_{i-1}a_i}\overset{\mathrm{SVD}}=U_{a_{i-1}p}^{\sigma_i} D_{pq}V_{qa_i}\quad\mathrm{[Right\;sweep]}
\end{equation}
and
\begin{equation}
A^{\sigma_i}_{a_{i-1}a_i}\overset{\mathrm{SVD}}=U_{a_{i-1}p} D_{pq}V_{qa_i}^{\sigma_i}\quad\mathrm{[Left\;sweep]}
\end{equation}

The algorithm as presented so far is when the expansion term is to be applied at full strength.  However, it is highly useful to decrease the effective magnitude of the expansion term such that the expansion is zero when the MPS is mostly converged.  This will aid convergence.  The algorithm presented in Ref.~\onlinecite{hubig2015strictly} uses two quantities to ensure that the value for the noise parameter $\gamma$ is set correctly.  The relationship stated in Ref.~\onlinecite{hubig2015strictly} is to keep the relationship $\Delta E_T\approx-0.3\Delta E_O$ where $\Delta E_T$ is the energy difference from the previous sweep and $\Delta E_O$ (both of which can be estimated from the energy values and truncation error) is the energy difference lost when truncating the tensor with the SVD.  In order to control the size of $\gamma$ further, a small decay can be multiplied onto the value at each step.

Because of the noise term, the 3S algorithm does not always converge the energy to the ground state of the system, but it generally gets very close.  Further, some bias in the wavefuncton based on the sweep direction occurs, just as for the original single-site methods.  The implementation here is that of the DMRjulia library \cite{dmrjulia1}.

The method then scales as $O(N_sdm^3)$, a reduction of a factor of $d$ from the two-site algorithm because only one physical index is present.

\end{appendix}

\newpage

\bibliography{refs}

\begin{thebibliography}{42}%
\makeatletter
\providecommand \@ifxundefined [1]{%
 \@ifx{#1\undefined}
}%
\providecommand \@ifnum [1]{%
 \ifnum #1\expandafter \@firstoftwo
 \else \expandafter \@secondoftwo
 \fi
}%
\providecommand \@ifx [1]{%
 \ifx #1\expandafter \@firstoftwo
 \else \expandafter \@secondoftwo
 \fi
}%
\providecommand \natexlab [1]{#1}%
\providecommand \enquote  [1]{``#1''}%
\providecommand \bibnamefont  [1]{#1}%
\providecommand \bibfnamefont [1]{#1}%
\providecommand \citenamefont [1]{#1}%
\providecommand \href@noop [0]{\@secondoftwo}%
\providecommand \href [0]{\begingroup \@sanitize@url \@href}%
\providecommand \@href[1]{\@@startlink{#1}\@@href}%
\providecommand \@@href[1]{\endgroup#1\@@endlink}%
\providecommand \@sanitize@url [0]{\catcode `\\12\catcode `\$12\catcode
  `\&12\catcode `\#12\catcode `\^12\catcode `\_12\catcode `\%12\relax}%
\providecommand \@@startlink[1]{}%
\providecommand \@@endlink[0]{}%
\providecommand \url  [0]{\begingroup\@sanitize@url \@url }%
\providecommand \@url [1]{\endgroup\@href {#1}{\urlprefix }}%
\providecommand \urlprefix  [0]{URL }%
\providecommand \Eprint [0]{\href }%
\providecommand \doibase [0]{http://dx.doi.org/}%
\providecommand \selectlanguage [0]{\@gobble}%
\providecommand \bibinfo  [0]{\@secondoftwo}%
\providecommand \bibfield  [0]{\@secondoftwo}%
\providecommand \translation [1]{[#1]}%
\providecommand \BibitemOpen [0]{}%
\providecommand \bibitemStop [0]{}%
\providecommand \bibitemNoStop [0]{.\EOS\space}%
\providecommand \EOS [0]{\spacefactor3000\relax}%
\providecommand \BibitemShut  [1]{\csname bibitem#1\endcsname}%
\let\auto@bib@innerbib\@empty
\bibitem [{\citenamefont {White}(1992)}]{white1992density}%
  \BibitemOpen
  \bibfield  {author} {\bibinfo {author} {\bibfnamefont {Steven~R}\
  \bibnamefont {White}},\ }\bibfield  {title} {\enquote {\bibinfo {title}
  {Density matrix formulation for quantum renormalization groups},}\
  }\href@noop {} {\bibfield  {journal} {\bibinfo  {journal} {Physical review
  letters}\ }\textbf {\bibinfo {volume} {69}},\ \bibinfo {pages} {2863}
  (\bibinfo {year} {1992})}\BibitemShut {NoStop}%
\bibitem [{\citenamefont {White}(1993)}]{white1993density}%
  \BibitemOpen
  \bibfield  {author} {\bibinfo {author} {\bibfnamefont {Steven~R}\
  \bibnamefont {White}},\ }\bibfield  {title} {\enquote {\bibinfo {title}
  {Density-matrix algorithms for quantum renormalization groups},}\ }\href@noop
  {} {\bibfield  {journal} {\bibinfo  {journal} {Physical Review B}\ }\textbf
  {\bibinfo {volume} {48}},\ \bibinfo {pages} {10345} (\bibinfo {year}
  {1993})}\BibitemShut {NoStop}%
\bibitem [{\citenamefont {Schollw{\"o}ck}(2005)}]{schollwock2005density}%
  \BibitemOpen
  \bibfield  {author} {\bibinfo {author} {\bibfnamefont {Ulrich}\ \bibnamefont
  {Schollw{\"o}ck}},\ }\bibfield  {title} {\enquote {\bibinfo {title} {The
  density-matrix renormalization group},}\ }\href@noop {} {\bibfield  {journal}
  {\bibinfo  {journal} {Reviews of modern physics}\ }\textbf {\bibinfo {volume}
  {77}},\ \bibinfo {pages} {259} (\bibinfo {year} {2005})}\BibitemShut
  {NoStop}%
\bibitem [{\citenamefont {Schollw{\"o}ck}(2011)}]{schollwock2011density}%
  \BibitemOpen
  \bibfield  {author} {\bibinfo {author} {\bibfnamefont {Ulrich}\ \bibnamefont
  {Schollw{\"o}ck}},\ }\bibfield  {title} {\enquote {\bibinfo {title} {The
  density-matrix renormalization group in the age of matrix product states},}\
  }\href@noop {} {\bibfield  {journal} {\bibinfo  {journal} {Annals of
  physics}\ }\textbf {\bibinfo {volume} {326}},\ \bibinfo {pages} {96--192}
  (\bibinfo {year} {2011})}\BibitemShut {NoStop}%
\bibitem [{\citenamefont {Stoudenmire}\ and\ \citenamefont
  {White}(2012)}]{stoudenmire2012studying}%
  \BibitemOpen
  \bibfield  {author} {\bibinfo {author} {\bibfnamefont {Edwin~M}\ \bibnamefont
  {Stoudenmire}}\ and\ \bibinfo {author} {\bibfnamefont {Steven~R}\
  \bibnamefont {White}},\ }\bibfield  {title} {\enquote {\bibinfo {title}
  {Studying two-dimensional systems with the density matrix renormalization
  group},}\ }\href@noop {} {\bibfield  {journal} {\bibinfo  {journal} {Annu.
  Rev. Condens. Matter Phys.}\ }\textbf {\bibinfo {volume} {3}},\ \bibinfo
  {pages} {111--128} (\bibinfo {year} {2012})}\BibitemShut {NoStop}%
\bibitem [{\citenamefont {{\"O}stlund}\ and\ \citenamefont
  {Rommer}(1995)}]{ostlund1995thermodynamic}%
  \BibitemOpen
  \bibfield  {author} {\bibinfo {author} {\bibfnamefont {Stellan}\ \bibnamefont
  {{\"O}stlund}}\ and\ \bibinfo {author} {\bibfnamefont {Stefan}\ \bibnamefont
  {Rommer}},\ }\bibfield  {title} {\enquote {\bibinfo {title} {Thermodynamic
  limit of density matrix renormalization},}\ }\href@noop {} {\bibfield
  {journal} {\bibinfo  {journal} {Physical review letters}\ }\textbf {\bibinfo
  {volume} {75}},\ \bibinfo {pages} {3537} (\bibinfo {year}
  {1995})}\BibitemShut {NoStop}%
\bibitem [{\citenamefont {Rommer}\ and\ \citenamefont
  {{\"O}stlund}(1997)}]{rommer1997class}%
  \BibitemOpen
  \bibfield  {author} {\bibinfo {author} {\bibfnamefont {Stefan}\ \bibnamefont
  {Rommer}}\ and\ \bibinfo {author} {\bibfnamefont {Stellan}\ \bibnamefont
  {{\"O}stlund}},\ }\bibfield  {title} {\enquote {\bibinfo {title} {Class of
  ansatz wave functions for one-dimensional spin systems and their relation to
  the density matrix renormalization group},}\ }\href@noop {} {\bibfield
  {journal} {\bibinfo  {journal} {Physical review b}\ }\textbf {\bibinfo
  {volume} {55}},\ \bibinfo {pages} {2164} (\bibinfo {year}
  {1997})}\BibitemShut {NoStop}%
\bibitem [{\citenamefont {Affleck}\ \emph {et~al.}(1988)\citenamefont
  {Affleck}, \citenamefont {Kennedy}, \citenamefont {Lieb},\ and\ \citenamefont
  {Tasaki}}]{affleck1988valence}%
  \BibitemOpen
  \bibfield  {author} {\bibinfo {author} {\bibfnamefont {Ian}\ \bibnamefont
  {Affleck}}, \bibinfo {author} {\bibfnamefont {Tom}\ \bibnamefont {Kennedy}},
  \bibinfo {author} {\bibfnamefont {Elliott~H}\ \bibnamefont {Lieb}}, \ and\
  \bibinfo {author} {\bibfnamefont {Hal}\ \bibnamefont {Tasaki}},\ }\bibfield
  {title} {\enquote {\bibinfo {title} {Valence bond ground states in isotropic
  quantum antiferromagnets},}\ }in\ \href@noop {} {\emph {\bibinfo {booktitle}
  {Condensed matter physics and exactly soluble models}}}\ (\bibinfo
  {publisher} {Springer},\ \bibinfo {year} {1988})\ pp.\ \bibinfo {pages}
  {253--304}\BibitemShut {NoStop}%
\bibitem [{\citenamefont {Feldt}\ and\ \citenamefont
  {Filippi}(2020)}]{feldt2020excited}%
  \BibitemOpen
  \bibfield  {author} {\bibinfo {author} {\bibfnamefont {Jonas}\ \bibnamefont
  {Feldt}}\ and\ \bibinfo {author} {\bibfnamefont {Claudia}\ \bibnamefont
  {Filippi}},\ }\bibfield  {title} {\enquote {\bibinfo {title} {Excited-state
  calculations with quantum monte carlo},}\ }\href@noop {} {\bibfield
  {journal} {\bibinfo  {journal} {arXiv preprint arXiv:2002.03622}\ } (\bibinfo
  {year} {2020})}\BibitemShut {NoStop}%
\bibitem [{\citenamefont {Vanderstraeten}\ \emph {et~al.}(2019)\citenamefont
  {Vanderstraeten}, \citenamefont {Haegeman},\ and\ \citenamefont
  {Verstraete}}]{vanderstraeten2019simulating}%
  \BibitemOpen
  \bibfield  {author} {\bibinfo {author} {\bibfnamefont {Laurens}\ \bibnamefont
  {Vanderstraeten}}, \bibinfo {author} {\bibfnamefont {Jutho}\ \bibnamefont
  {Haegeman}}, \ and\ \bibinfo {author} {\bibfnamefont {Frank}\ \bibnamefont
  {Verstraete}},\ }\bibfield  {title} {\enquote {\bibinfo {title} {Simulating
  excitation spectra with projected entangled-pair states},}\ }\href@noop {}
  {\bibfield  {journal} {\bibinfo  {journal} {Physical Review B}\ }\textbf
  {\bibinfo {volume} {99}},\ \bibinfo {pages} {165121} (\bibinfo {year}
  {2019})}\BibitemShut {NoStop}%
\bibitem [{\citenamefont {D'Azevedo}\ \emph {et~al.}(2019)\citenamefont
  {D'Azevedo}, \citenamefont {Elwasif}, \citenamefont {Patel},\ and\
  \citenamefont {Alvarez}}]{d2019targeting}%
  \BibitemOpen
  \bibfield  {author} {\bibinfo {author} {\bibfnamefont {EF}~\bibnamefont
  {D'Azevedo}}, \bibinfo {author} {\bibfnamefont {WR}~\bibnamefont {Elwasif}},
  \bibinfo {author} {\bibfnamefont {ND}~\bibnamefont {Patel}}, \ and\ \bibinfo
  {author} {\bibfnamefont {G}~\bibnamefont {Alvarez}},\ }\bibfield  {title}
  {\enquote {\bibinfo {title} {Targeting multiple states in the density matrix
  renormalization group with the singular value decomposition},}\ }\href@noop
  {} {\bibfield  {journal} {\bibinfo  {journal} {arXiv preprint
  arXiv:1902.09621}\ } (\bibinfo {year} {2019})}\BibitemShut {NoStop}%
\bibitem [{\citenamefont {Ponsioen}\ and\ \citenamefont
  {Corboz}(2020)}]{ponsioen2020excitations}%
  \BibitemOpen
  \bibfield  {author} {\bibinfo {author} {\bibfnamefont {Boris}\ \bibnamefont
  {Ponsioen}}\ and\ \bibinfo {author} {\bibfnamefont {Philippe}\ \bibnamefont
  {Corboz}},\ }\bibfield  {title} {\enquote {\bibinfo {title} {Excitations with
  projected entangled pair states using the corner transfer matrix method},}\
  }\href@noop {} {\bibfield  {journal} {\bibinfo  {journal} {Physical Review
  B}\ }\textbf {\bibinfo {volume} {101}},\ \bibinfo {pages} {195109} (\bibinfo
  {year} {2020})}\BibitemShut {NoStop}%
\bibitem [{\citenamefont {Boschi}\ and\ \citenamefont
  {Ortolani}(2004)}]{boschi2004investigation}%
  \BibitemOpen
  \bibfield  {author} {\bibinfo {author} {\bibfnamefont {C~Degli~Esposti}\
  \bibnamefont {Boschi}}\ and\ \bibinfo {author} {\bibfnamefont {Fabio}\
  \bibnamefont {Ortolani}},\ }\bibfield  {title} {\enquote {\bibinfo {title}
  {Investigation of quantum phase transitions using multi-target dmrg
  methods},}\ }\href@noop {} {\bibfield  {journal} {\bibinfo  {journal} {The
  European Physical Journal B-Condensed Matter and Complex Systems}\ }\textbf
  {\bibinfo {volume} {41}},\ \bibinfo {pages} {503--516} (\bibinfo {year}
  {2004})}\BibitemShut {NoStop}%
\bibitem [{\citenamefont {Hallberg}(1995)}]{hallberg1995density}%
  \BibitemOpen
  \bibfield  {author} {\bibinfo {author} {\bibfnamefont {Karen~A}\ \bibnamefont
  {Hallberg}},\ }\bibfield  {title} {\enquote {\bibinfo {title} {Density-matrix
  algorithm for the calculation of dynamical properties of low-dimensional
  systems},}\ }\href@noop {} {\bibfield  {journal} {\bibinfo  {journal}
  {Physical Review B}\ }\textbf {\bibinfo {volume} {52}},\ \bibinfo {pages}
  {R9827} (\bibinfo {year} {1995})}\BibitemShut {NoStop}%
\bibitem [{\citenamefont {Jeckelmann}(2002)}]{jeckelmann2002dynamical}%
  \BibitemOpen
  \bibfield  {author} {\bibinfo {author} {\bibfnamefont {Eric}\ \bibnamefont
  {Jeckelmann}},\ }\bibfield  {title} {\enquote {\bibinfo {title} {Dynamical
  density-matrix renormalization-group method},}\ }\href@noop {} {\bibfield
  {journal} {\bibinfo  {journal} {Physical Review B}\ }\textbf {\bibinfo
  {volume} {66}},\ \bibinfo {pages} {045114} (\bibinfo {year}
  {2002})}\BibitemShut {NoStop}%
\bibitem [{\citenamefont {Chepiga}\ and\ \citenamefont
  {Mila}(2017)}]{chepiga2017excitation}%
  \BibitemOpen
  \bibfield  {author} {\bibinfo {author} {\bibfnamefont {Natalia}\ \bibnamefont
  {Chepiga}}\ and\ \bibinfo {author} {\bibfnamefont {Fr{\'e}d{\'e}ric}\
  \bibnamefont {Mila}},\ }\bibfield  {title} {\enquote {\bibinfo {title}
  {Excitation spectrum and density matrix renormalization group iterations},}\
  }\href@noop {} {\bibfield  {journal} {\bibinfo  {journal} {Physical Review
  B}\ }\textbf {\bibinfo {volume} {96}},\ \bibinfo {pages} {054425} (\bibinfo
  {year} {2017})}\BibitemShut {NoStop}%
\bibitem [{\citenamefont {Van~Damme}\ \emph {et~al.}(2021)\citenamefont
  {Van~Damme}, \citenamefont {Vanhove}, \citenamefont {Haegeman}, \citenamefont
  {Verstraete},\ and\ \citenamefont {Vanderstraeten}}]{van2021efficient}%
  \BibitemOpen
  \bibfield  {author} {\bibinfo {author} {\bibfnamefont {Maarten}\ \bibnamefont
  {Van~Damme}}, \bibinfo {author} {\bibfnamefont {Robijn}\ \bibnamefont
  {Vanhove}}, \bibinfo {author} {\bibfnamefont {Jutho}\ \bibnamefont
  {Haegeman}}, \bibinfo {author} {\bibfnamefont {Frank}\ \bibnamefont
  {Verstraete}}, \ and\ \bibinfo {author} {\bibfnamefont {Laurens}\
  \bibnamefont {Vanderstraeten}},\ }\bibfield  {title} {\enquote {\bibinfo
  {title} {Efficient matrix product state methods for extracting spectral
  information on rings and cylinders},}\ }\href@noop {} {\bibfield  {journal}
  {\bibinfo  {journal} {Physical Review B}\ }\textbf {\bibinfo {volume}
  {104}},\ \bibinfo {pages} {115142} (\bibinfo {year} {2021})}\BibitemShut
  {NoStop}%
\bibitem [{\citenamefont {Khemani}\ \emph {et~al.}(2016)\citenamefont
  {Khemani}, \citenamefont {Pollmann},\ and\ \citenamefont
  {Sondhi}}]{khemani2016obtaining}%
  \BibitemOpen
  \bibfield  {author} {\bibinfo {author} {\bibfnamefont {Vedika}\ \bibnamefont
  {Khemani}}, \bibinfo {author} {\bibfnamefont {Frank}\ \bibnamefont
  {Pollmann}}, \ and\ \bibinfo {author} {\bibfnamefont {Shivaji~Lal}\
  \bibnamefont {Sondhi}},\ }\bibfield  {title} {\enquote {\bibinfo {title}
  {Obtaining highly excited eigenstates of many-body localized hamiltonians by
  the density matrix renormalization group approach},}\ }\href@noop {}
  {\bibfield  {journal} {\bibinfo  {journal} {Physical review letters}\
  }\textbf {\bibinfo {volume} {116}},\ \bibinfo {pages} {247204} (\bibinfo
  {year} {2016})}\BibitemShut {NoStop}%
\bibitem [{\citenamefont {Yu}\ \emph {et~al.}(2017)\citenamefont {Yu},
  \citenamefont {Pekker},\ and\ \citenamefont {Clark}}]{yu2017finding}%
  \BibitemOpen
  \bibfield  {author} {\bibinfo {author} {\bibfnamefont {Xiongjie}\
  \bibnamefont {Yu}}, \bibinfo {author} {\bibfnamefont {David}\ \bibnamefont
  {Pekker}}, \ and\ \bibinfo {author} {\bibfnamefont {Bryan~K}\ \bibnamefont
  {Clark}},\ }\bibfield  {title} {\enquote {\bibinfo {title} {Finding matrix
  product state representations of highly excited eigenstates of many-body
  localized hamiltonians},}\ }\href@noop {} {\bibfield  {journal} {\bibinfo
  {journal} {Physical review letters}\ }\textbf {\bibinfo {volume} {118}},\
  \bibinfo {pages} {017201} (\bibinfo {year} {2017})}\BibitemShut {NoStop}%
\bibitem [{\citenamefont {Baker}\ \emph {et~al.}(2021)\citenamefont {Baker},
  \citenamefont {Desrosiers}, \citenamefont {Tremblay},\ and\ \citenamefont
  {Thompson}}]{bakerCJP21}%
  \BibitemOpen
  \bibfield  {author} {\bibinfo {author} {\bibfnamefont {Thomas~E}\
  \bibnamefont {Baker}}, \bibinfo {author} {\bibfnamefont {Samuel}\
  \bibnamefont {Desrosiers}}, \bibinfo {author} {\bibfnamefont {Maxime}\
  \bibnamefont {Tremblay}}, \ and\ \bibinfo {author} {\bibfnamefont {Martin~P}\
  \bibnamefont {Thompson}},\ }\bibfield  {title} {\enquote {\bibinfo {title}
  {{M{\'e}thodes de calcul avec r{\'e}seaux de tenseurs en physique}},}\ }\href
  {\doibase https://doi.org/10.1139/cjp-2019-0611} {\bibfield  {journal}
  {\bibinfo  {journal} {Canadian Journal of Physics}\ }\textbf {\bibinfo
  {volume} {99}},\ \bibinfo {pages} {4} (\bibinfo {year} {2021})}\BibitemShut
  {NoStop}%
\bibitem [{\citenamefont {Baker}\ \emph {et~al.}(2019)\citenamefont {Baker},
  \citenamefont {Desrosiers}, \citenamefont {Tremblay},\ and\ \citenamefont
  {Thompson}}]{baker2019m}%
  \BibitemOpen
  \bibfield  {author} {\bibinfo {author} {\bibfnamefont {Thomas~E}\
  \bibnamefont {Baker}}, \bibinfo {author} {\bibfnamefont {Samuel}\
  \bibnamefont {Desrosiers}}, \bibinfo {author} {\bibfnamefont {Maxime}\
  \bibnamefont {Tremblay}}, \ and\ \bibinfo {author} {\bibfnamefont {Martin~P}\
  \bibnamefont {Thompson}},\ }\bibfield  {title} {\enquote {\bibinfo {title}
  {Basic tensor network computations in physics},}\ }\href@noop {} {\bibfield
  {journal} {\bibinfo  {journal} {arXiv preprint arXiv:1911.11566, p.~19}\ }
  (\bibinfo {year} {2019})}\BibitemShut {NoStop}%
\bibitem [{\citenamefont {Huang}\ \emph {et~al.}(2018)\citenamefont {Huang},
  \citenamefont {Liao}, \citenamefont {Liu}, \citenamefont {Xie}, \citenamefont
  {Xie}, \citenamefont {Zhao}, \citenamefont {Chen},\ and\ \citenamefont
  {Xiang}}]{huang2018generalized}%
  \BibitemOpen
  \bibfield  {author} {\bibinfo {author} {\bibfnamefont {Rui-Zhen}\
  \bibnamefont {Huang}}, \bibinfo {author} {\bibfnamefont {Hai-Jun}\
  \bibnamefont {Liao}}, \bibinfo {author} {\bibfnamefont {Zhi-Yuan}\
  \bibnamefont {Liu}}, \bibinfo {author} {\bibfnamefont {Hai-Dong}\
  \bibnamefont {Xie}}, \bibinfo {author} {\bibfnamefont {Zhi-Yuan}\
  \bibnamefont {Xie}}, \bibinfo {author} {\bibfnamefont {Hui-Hai}\ \bibnamefont
  {Zhao}}, \bibinfo {author} {\bibfnamefont {Jing}\ \bibnamefont {Chen}}, \
  and\ \bibinfo {author} {\bibfnamefont {Tao}\ \bibnamefont {Xiang}},\
  }\bibfield  {title} {\enquote {\bibinfo {title} {Generalized lanczos method
  for systematic optimization of tensor network states},}\ }\href@noop {}
  {\bibfield  {journal} {\bibinfo  {journal} {Chinese Physics B}\ }\textbf
  {\bibinfo {volume} {27}},\ \bibinfo {pages} {070501} (\bibinfo {year}
  {2018})}\BibitemShut {NoStop}%
\bibitem [{\citenamefont {Cullum}\ and\ \citenamefont
  {Donath}(1974)}]{cullum1974block}%
  \BibitemOpen
  \bibfield  {author} {\bibinfo {author} {\bibfnamefont {Jane}\ \bibnamefont
  {Cullum}}\ and\ \bibinfo {author} {\bibfnamefont {William~E}\ \bibnamefont
  {Donath}},\ }\bibfield  {title} {\enquote {\bibinfo {title} {A block lanczos
  algorithm for computing the q algebraically largest eigenvalues and a
  corresponding eigenspace of large, sparse, real symmetric matrices},}\ }in\
  \href@noop {} {\emph {\bibinfo {booktitle} {1974 IEEE Conference on Decision
  and Control including the 13th Symposium on Adaptive Processes}}}\ (\bibinfo
  {organization} {IEEE},\ \bibinfo {year} {1974})\ pp.\ \bibinfo {pages}
  {505--509}\BibitemShut {NoStop}%
\bibitem [{\citenamefont {Bai}\ \emph {et~al.}(2000)\citenamefont {Bai},
  \citenamefont {Demmel}, \citenamefont {Dongarra}, \citenamefont {Ruhe},\ and\
  \citenamefont {van~der Vorst}}]{bai2000templates}%
  \BibitemOpen
  \bibfield  {author} {\bibinfo {author} {\bibfnamefont {Zhaojun}\ \bibnamefont
  {Bai}}, \bibinfo {author} {\bibfnamefont {James}\ \bibnamefont {Demmel}},
  \bibinfo {author} {\bibfnamefont {Jack}\ \bibnamefont {Dongarra}}, \bibinfo
  {author} {\bibfnamefont {Axel}\ \bibnamefont {Ruhe}}, \ and\ \bibinfo
  {author} {\bibfnamefont {Henk}\ \bibnamefont {van~der Vorst}},\ }\href@noop
  {} {\emph {\bibinfo {title} {Templates for the solution of algebraic
  eigenvalue problems: a practical guide}}}\ (\bibinfo  {publisher} {SIAM},\
  \bibinfo {year} {2000})\BibitemShut {NoStop}%
\bibitem [{\citenamefont {S{\'e}n{\'e}chal}(2008)}]{senechal2008introduction}%
  \BibitemOpen
  \bibfield  {author} {\bibinfo {author} {\bibfnamefont {David}\ \bibnamefont
  {S{\'e}n{\'e}chal}},\ }\bibfield  {title} {\enquote {\bibinfo {title} {An
  introduction to quantum cluster methods},}\ }\href@noop {} {\bibfield
  {journal} {\bibinfo  {journal} {arXiv preprint arXiv:0806.2690}\ } (\bibinfo
  {year} {2008})}\BibitemShut {NoStop}%
\bibitem [{\citenamefont {Di~Paolo}\ \emph {et~al.}(2021)\citenamefont
  {Di~Paolo}, \citenamefont {Baker}, \citenamefont {Foley}, \citenamefont
  {S{\'e}n{\'e}chal},\ and\ \citenamefont {Blais}}]{di2021efficient}%
  \BibitemOpen
  \bibfield  {author} {\bibinfo {author} {\bibfnamefont {Agustin}\ \bibnamefont
  {Di~Paolo}}, \bibinfo {author} {\bibfnamefont {Thomas~E}\ \bibnamefont
  {Baker}}, \bibinfo {author} {\bibfnamefont {Alexandre}\ \bibnamefont
  {Foley}}, \bibinfo {author} {\bibfnamefont {David}\ \bibnamefont
  {S{\'e}n{\'e}chal}}, \ and\ \bibinfo {author} {\bibfnamefont {Alexandre}\
  \bibnamefont {Blais}},\ }\bibfield  {title} {\enquote {\bibinfo {title}
  {Efficient modeling of superconducting quantum circuits with tensor
  networks},}\ }\href {\doibase https://doi.org/10.1038/s41534-020-00352-4}
  {\bibfield  {journal} {\bibinfo  {journal} {npj Quantum Information}\
  }\textbf {\bibinfo {volume} {7}},\ \bibinfo {pages} {1--11} (\bibinfo {year}
  {2021})}\BibitemShut {NoStop}%
\bibitem [{\citenamefont {Baker}\ and\ \citenamefont
  {Thompson}(2021)}]{dmrjulia1}%
  \BibitemOpen
  \bibfield  {author} {\bibinfo {author} {\bibfnamefont {Thomas~E}\
  \bibnamefont {Baker}}\ and\ \bibinfo {author} {\bibfnamefont {Martin~P}\
  \bibnamefont {Thompson}},\ }\bibfield  {title} {\enquote {\bibinfo {title}
  {{DMRjulia I. Basic construction of a tensor network library for the density
  matrix renormalization group}},}\ }\href@noop {} {\bibfield  {journal}
  {\bibinfo  {journal} {arXiv preprint arXiv:2109.03120}\ } (\bibinfo {year}
  {2021})}\BibitemShut {NoStop}%
\bibitem [{\citenamefont {Bhatia}(1997)}]{bhatia1997matrix}%
  \BibitemOpen
  \bibfield  {author} {\bibinfo {author} {\bibfnamefont {R}~\bibnamefont
  {Bhatia}},\ }\href {\doibase 10.1007/978-1-4612-0653-8} {\emph {\bibinfo
  {title} {Matrix Analysis}}}\ (\bibinfo  {publisher} {Springer-Verlag New
  York},\ \bibinfo {year} {1997})\BibitemShut {NoStop}%
\bibitem [{\citenamefont {Hauschild}\ and\ \citenamefont
  {Pollmann}(2018)}]{hauschild2018efficient}%
  \BibitemOpen
  \bibfield  {author} {\bibinfo {author} {\bibfnamefont {Johannes}\
  \bibnamefont {Hauschild}}\ and\ \bibinfo {author} {\bibfnamefont {Frank}\
  \bibnamefont {Pollmann}},\ }\bibfield  {title} {\enquote {\bibinfo {title}
  {Efficient numerical simulations with tensor networks: Tensor network python
  (tenpy)},}\ }\href@noop {} {\bibfield  {journal} {\bibinfo  {journal}
  {SciPost Physics Lecture Notes}\ ,\ \bibinfo {pages} {005}} (\bibinfo {year}
  {2018})}\BibitemShut {NoStop}%
\bibitem [{\citenamefont {Or{\'u}s}(2019)}]{orus2019tensor}%
  \BibitemOpen
  \bibfield  {author} {\bibinfo {author} {\bibfnamefont {Rom{\'a}n}\
  \bibnamefont {Or{\'u}s}},\ }\bibfield  {title} {\enquote {\bibinfo {title}
  {Tensor networks for complex quantum systems},}\ }\href@noop {} {\bibfield
  {journal} {\bibinfo  {journal} {Nature Reviews Physics}\ }\textbf {\bibinfo
  {volume} {1}},\ \bibinfo {pages} {538--550} (\bibinfo {year}
  {2019})}\BibitemShut {NoStop}%
\bibitem [{\citenamefont {McCulloch}(2007)}]{mcculloch2007density}%
  \BibitemOpen
  \bibfield  {author} {\bibinfo {author} {\bibfnamefont {Ian~P}\ \bibnamefont
  {McCulloch}},\ }\bibfield  {title} {\enquote {\bibinfo {title} {From
  density-matrix renormalization group to matrix product states},}\ }\href@noop
  {} {\bibfield  {journal} {\bibinfo  {journal} {Journal of Statistical
  Mechanics: Theory and Experiment}\ }\textbf {\bibinfo {volume} {2007}},\
  \bibinfo {pages} {P10014} (\bibinfo {year} {2007})}\BibitemShut {NoStop}%
\bibitem [{\citenamefont {Hubig}\ \emph {et~al.}(2015)\citenamefont {Hubig},
  \citenamefont {McCulloch}, \citenamefont {Schollw{\"o}ck},\ and\
  \citenamefont {Wolf}}]{hubig2015strictly}%
  \BibitemOpen
  \bibfield  {author} {\bibinfo {author} {\bibfnamefont {Claudius}\
  \bibnamefont {Hubig}}, \bibinfo {author} {\bibfnamefont {Ian~P}\ \bibnamefont
  {McCulloch}}, \bibinfo {author} {\bibfnamefont {Ulrich}\ \bibnamefont
  {Schollw{\"o}ck}}, \ and\ \bibinfo {author} {\bibfnamefont {F~Alexander}\
  \bibnamefont {Wolf}},\ }\bibfield  {title} {\enquote {\bibinfo {title}
  {Strictly single-site dmrg algorithm with subspace expansion},}\ }\href@noop
  {} {\bibfield  {journal} {\bibinfo  {journal} {Physical Review B}\ }\textbf
  {\bibinfo {volume} {91}},\ \bibinfo {pages} {155115} (\bibinfo {year}
  {2015})}\BibitemShut {NoStop}%
\bibitem [{\citenamefont {Haldane}(1983{\natexlab{a}})}]{haldane1983continuum}%
  \BibitemOpen
  \bibfield  {author} {\bibinfo {author} {\bibfnamefont {F~Duncan~M}\
  \bibnamefont {Haldane}},\ }\bibfield  {title} {\enquote {\bibinfo {title}
  {Continuum dynamics of the 1-d heisenberg antiferromagnet: Identification
  with the o (3) nonlinear sigma model},}\ }\href@noop {} {\bibfield  {journal}
  {\bibinfo  {journal} {Physics letters a}\ }\textbf {\bibinfo {volume} {93}},\
  \bibinfo {pages} {464--468} (\bibinfo {year}
  {1983}{\natexlab{a}})}\BibitemShut {NoStop}%
\bibitem [{\citenamefont {Haldane}(1983{\natexlab{b}})}]{haldane1983nonlinear}%
  \BibitemOpen
  \bibfield  {author} {\bibinfo {author} {\bibfnamefont {F~Duncan~M}\
  \bibnamefont {Haldane}},\ }\bibfield  {title} {\enquote {\bibinfo {title}
  {Nonlinear field theory of large-spin heisenberg antiferromagnets:
  semiclassically quantized solitons of the one-dimensional easy-axis n{\'e}el
  state},}\ }\href@noop {} {\bibfield  {journal} {\bibinfo  {journal} {Physical
  review letters}\ }\textbf {\bibinfo {volume} {50}},\ \bibinfo {pages} {1153}
  (\bibinfo {year} {1983}{\natexlab{b}})}\BibitemShut {NoStop}%
\bibitem [{\citenamefont {Lieb}(1989)}]{lieb1989two}%
  \BibitemOpen
  \bibfield  {author} {\bibinfo {author} {\bibfnamefont {Elliott~H}\
  \bibnamefont {Lieb}},\ }\bibfield  {title} {\enquote {\bibinfo {title} {Two
  theorems on the hubbard model},}\ }\href@noop {} {\bibfield  {journal}
  {\bibinfo  {journal} {Physical review letters}\ }\textbf {\bibinfo {volume}
  {62}},\ \bibinfo {pages} {1201} (\bibinfo {year} {1989})}\BibitemShut
  {NoStop}%
\bibitem [{\citenamefont {Kramers}(1930)}]{kramers1930theorie}%
  \BibitemOpen
  \bibfield  {author} {\bibinfo {author} {\bibfnamefont {Hendrik~Antoon}\
  \bibnamefont {Kramers}},\ }\bibfield  {title} {\enquote {\bibinfo {title}
  {Th{\'e}orie g{\'e}n{\'e}rale de la rotation paramagn{\'e}tique dans les
  cristaux},}\ }\href@noop {} {\bibfield  {journal} {\bibinfo  {journal} {Proc.
  Acad. Amst}\ }\textbf {\bibinfo {volume} {33}} (\bibinfo {year}
  {1930})}\BibitemShut {NoStop}%
\bibitem [{\citenamefont {Baker}\ \emph {et~al.}(2015)\citenamefont {Baker},
  \citenamefont {Stoudenmire}, \citenamefont {Wagner}, \citenamefont {Burke},\
  and\ \citenamefont {White}}]{bakerPRB15}%
  \BibitemOpen
  \bibfield  {author} {\bibinfo {author} {\bibfnamefont {Thomas~E}\
  \bibnamefont {Baker}}, \bibinfo {author} {\bibfnamefont {E~Miles}\
  \bibnamefont {Stoudenmire}}, \bibinfo {author} {\bibfnamefont {Lucas~O}\
  \bibnamefont {Wagner}}, \bibinfo {author} {\bibfnamefont {Kieron}\
  \bibnamefont {Burke}}, \ and\ \bibinfo {author} {\bibfnamefont {Steven~R}\
  \bibnamefont {White}},\ }\bibfield  {title} {\enquote {\bibinfo {title}
  {{One-dimensional mimicking of electronic structure: The case for
  exponentials}},}\ }\href {\doibase
  https://doi.org/10.1103/PhysRevB.91.235141} {\bibfield  {journal} {\bibinfo
  {journal} {Phys.~Rev.~B}\ }\textbf {\bibinfo {volume} {91}},\ \bibinfo
  {pages} {235141} (\bibinfo {year} {2015})}\BibitemShut {NoStop}%
\bibitem [{\citenamefont {Wagner}\ \emph {et~al.}(2014)\citenamefont {Wagner},
  \citenamefont {Baker}, \citenamefont {Stoudenmire}, \citenamefont {Burke},\
  and\ \citenamefont {White}}]{wagnerPRB14}%
  \BibitemOpen
  \bibfield  {author} {\bibinfo {author} {\bibfnamefont {Lucas~O}\ \bibnamefont
  {Wagner}}, \bibinfo {author} {\bibfnamefont {Thomas~E}\ \bibnamefont
  {Baker}}, \bibinfo {author} {\bibfnamefont {EM}~\bibnamefont {Stoudenmire}},
  \bibinfo {author} {\bibfnamefont {Kieron}\ \bibnamefont {Burke}}, \ and\
  \bibinfo {author} {\bibfnamefont {Steven~R}\ \bibnamefont {White}},\
  }\bibfield  {title} {\enquote {\bibinfo {title} {{Kohn-Sham calculations with
  the exact functional}},}\ }\href {\doibase
  https://doi.org/10.1103/PhysRevB.90.045109} {\bibfield  {journal} {\bibinfo
  {journal} {Phys.~Rev.~B}\ }\textbf {\bibinfo {volume} {90}},\ \bibinfo
  {pages} {045109} (\bibinfo {year} {2014})}\BibitemShut {NoStop}%
\bibitem [{\citenamefont {Dolgov}\ \emph {et~al.}(2014)\citenamefont {Dolgov},
  \citenamefont {Khoromskij}, \citenamefont {Oseledets},\ and\ \citenamefont
  {Savostyanov}}]{dolgov2014computation}%
  \BibitemOpen
  \bibfield  {author} {\bibinfo {author} {\bibfnamefont {Sergey~V}\
  \bibnamefont {Dolgov}}, \bibinfo {author} {\bibfnamefont {Boris~N}\
  \bibnamefont {Khoromskij}}, \bibinfo {author} {\bibfnamefont {Ivan~V}\
  \bibnamefont {Oseledets}}, \ and\ \bibinfo {author} {\bibfnamefont
  {Dmitry~V}\ \bibnamefont {Savostyanov}},\ }\bibfield  {title} {\enquote
  {\bibinfo {title} {Computation of extreme eigenvalues in higher dimensions
  using block tensor train format},}\ }\href@noop {} {\bibfield  {journal}
  {\bibinfo  {journal} {Computer Physics Communications}\ }\textbf {\bibinfo
  {volume} {185}},\ \bibinfo {pages} {1207--1216} (\bibinfo {year}
  {2014})}\BibitemShut {NoStop}%
\bibitem [{\citenamefont {Knyazev}(2001)}]{knyazev2001toward}%
  \BibitemOpen
  \bibfield  {author} {\bibinfo {author} {\bibfnamefont {Andrew~V}\
  \bibnamefont {Knyazev}},\ }\bibfield  {title} {\enquote {\bibinfo {title}
  {Toward the optimal preconditioned eigensolver: Locally optimal block
  preconditioned conjugate gradient method},}\ }\href@noop {} {\bibfield
  {journal} {\bibinfo  {journal} {SIAM journal on scientific computing}\
  }\textbf {\bibinfo {volume} {23}},\ \bibinfo {pages} {517--541} (\bibinfo
  {year} {2001})}\BibitemShut {NoStop}%
\bibitem [{\citenamefont {Baker}()}]{dmrjulia}%
  \BibitemOpen
  \bibfield  {author} {\bibinfo {author} {\bibfnamefont {Thomas~E.}\
  \bibnamefont {Baker}},\ }\href@noop {} {\enquote {\bibinfo {title}
  {{DMRjulia}},}\ }\bibinfo {howpublished}
  {\url{https://github.com/bakerte/DMRJtensor.jl}}\BibitemShut {NoStop}%
\bibitem [{\citenamefont {Baker}(2021)}]{baker2021lanczos}%
  \BibitemOpen
  \bibfield  {author} {\bibinfo {author} {\bibfnamefont {Thomas~E}\
  \bibnamefont {Baker}},\ }\bibfield  {title} {\enquote {\bibinfo {title}
  {{Lanczos recursion on a quantum computer for the Green's function and ground
  state}},}\ }\href {\doibase https://doi.org/10.1103/PhysRevA.103.032404}
  {\bibfield  {journal} {\bibinfo  {journal} {Physical Review A}\ }\textbf
  {\bibinfo {volume} {103}},\ \bibinfo {pages} {032404} (\bibinfo {year}
  {2021})}\BibitemShut {NoStop}%
\end{thebibliography}%

\end{document}